\documentstyle[epsfig]{mn}

\newif\ifAMStwofonts

\def\lesssim{\mathrel{\hbox{\rlap{\hbox{\lower4pt\hbox{$\sim$}}}\hbox{$<$}}}}
\def\gtrsim{\mathrel{\hbox{\rlap{\hbox{\lower4pt\hbox{$\sim$}}}\hbox{$>$}}}}

\ifoldfss
  \ifCUPmtlplainloaded \else
    \NewTextAlphabet{textbfit} {cmbxti10} {}
    \NewTextAlphabet{textbfss} {cmssbx10} {}
    \NewMathAlphabet{mathbfit} {cmbxti10} {} 
    \NewMathAlphabet{mathbfss} {cmssbx10} {} 
  \fi
  \ifAMStwofonts
    \ifCUPmtlplainloaded \else
      \NewSymbolFont{upmath} {eurm10}
      \NewSymbolFont{AMSa} {msam10}
      \NewMathSymbol{\upi}     {0}{upmath}{19}
      \NewMathSymbol{\umu}     {0}{upmath}{16}
      \NewMathSymbol{\upartial}{0}{upmath}{40}
      \NewMathSymbol{\leqslant}{3}{AMSa}{36}
      \NewMathSymbol{\geqslant}{3}{AMSa}{3E}

       \let\le=\leqslant
       \let\ge=\geqslant
    \fi
  \fi
\fi 

\ifnfssone
  \newmathalphabet{\mathit}
  \addtoversion{normal}{\mathit}{cmr}{m}{it}
  \addtoversion{bold}{\mathit}{cmr}{bx}{it}
  \newmathalphabet{\mathbfit} 
  \addtoversion{normal}{\mathbfit}{cmr}{bx}{it}
  \addtoversion{bold}{\mathbfit}{cmr}{bx}{it}
  \newmathalphabet{\mathbfss} 
  \addtoversion{normal}{\mathbfss}{cmss}{bx}{n}
  \addtoversion{bold}{\mathbfss}{cmss}{bx}{n}
  \ifAMStwofonts
    \ifCUPmtlplainloaded \else
      \UseAMStwoboldmath
      \makeatletter
      \new@mathgroup\upmath@group
      \define@mathgroup\mv@normal\upmath@group{eur}{m}{n}
      \define@mathgroup\mv@bold\upmath@group{eur}{b}{n}
      \edef\UPM{\hexnumber\upmath@group}
      \new@mathgroup\amsa@group
      \define@mathgroup\mv@normal\amsa@group{msa}{m}{n}
      \define@mathgroup\mv@bold\amsa@group{msa}{m}{n}
      \edef\AMSa{\hexnumber\amsa@group}
      \makeatother
      \mathchardef\upi="0\UPM19
      \mathchardef\umu="0\UPM16
      \mathchardef\upartial="0\UPM40
      \mathchardef\leqslant="3\AMSa36
      \mathchardef\geqslant="3\AMSa3E

       \let\le=\leqslant
       \let\ge=\geqslant
    \fi
  \fi
\fi 

\ifnfsstwo
  \DeclareMathAlphabet{\mathbfit}{OT1}{cmr}{bx}{it}
  \SetMathAlphabet\mathbfit{bold}{OT1}{cmr}{bx}{it}
  \DeclareMathAlphabet{\mathbfss}{OT1}{cmss}{bx}{n}
  \SetMathAlphabet\mathbfss{bold}{OT1}{cmss}{bx}{n}
  \ifAMStwofonts
    \ifCUPmtlplainloaded \else
      \DeclareSymbolFont{UPM}{U}{eur}{m}{n}
      \SetSymbolFont{UPM}{bold}{U}{eur}{b}{n}
      \DeclareSymbolFont{AMSa}{U}{msa}{m}{n}
      \DeclareMathSymbol{\upi}{0}{UPM}{"19}
      \DeclareMathSymbol{\umu}{0}{UPM}{"16}
      \DeclareMathSymbol{\upartial}{0}{UPM}{"40}
      \DeclareMathSymbol{\leqslant}{3}{AMSa}{"36}
      \DeclareMathSymbol{\geqslant}{3}{AMSa}{"3E}

       \let\le=\leqslant
       \let\ge=\geqslant
    \fi
  \fi
\fi 

\ifCUPmtlplainloaded \else
  \ifAMStwofonts \else 
    \def\upi{\pi}
    \def\umu{\mu}
    \def\upartial{\partial}
  \fi
\fi

\title[Probing Early Star Formation through Radio to IR Absorption Lines in Very High-z GRBs]
{
The Radio to Infrared Emission of Very High Redshift Gamma-Ray Bursts:
Probing Early Star Formation through Molecular and Atomic Absorption Lines
}
\author[Susumu Inoue, Kazuyuki Omukai \& Benedetta Ciardi]
       {Susumu Inoue$^{1,2,3}$,
        Kazuyuki Omukai$^{3}$ and Benedetta Ciardi$^{2}$\\
        $^1$Max-Planck-Institut f\"ur Kernphysik, Postfach 103980, 69029 Heidelberg, Germany\\
        $^2$Max-Planck-Institut f\"ur Astrophysik, Karl-Schwarzschild-Str. 1, Postfach 1317, 85741 Garching, Germany\\
        $^3$National Astronomical Observatory of Japan, 2-21-1 Osawa, Mitaka, Tokyo, Japan 181-8588}
\date{Submitted to MNRAS}

\pagerange{\pageref{firstpage}--\pageref{lastpage}}
\pubyear{2004}

\begin{document}

\maketitle

\label{firstpage}

\begin{abstract}
We evaluate the broadband afterglow emission of very high redshift gamma-ray bursts (GRBs) 
using standard relativistic blastwave models with both forward and reverse shock components.
For a broad range of parameters,
a generic property for GRBs at redshifts $z \sim$ 5--30
is that the emission peaks in the millimeter to far-infrared bands with milli-Jansky flux levels,
first at a few hours after the burst due to the reverse shock,
and then again for several days afterwards with somewhat lower flux due to the forward shock.
The radio, submillimeter and infrared continuum emission should be readily detectable out to $z \ga 30$
by the Atacama Large Millimeter Array (ALMA), Extended Very Large Array (EVLA),
Square Kilometer Array (SKA) and other facilities.
For relatively bright bursts, spectroscopic measurements
of molecular and atomic absorption lines due to ambient protostellar gas may be possible.
Utilizing models of primordial protostellar clouds,
we show that under certain conditions, appreciable absorption
may be caused by HD rotational transitions even in metal-free environments.
After sufficient metal enrichment,
absorption from CO rotational transitions
and [OI] fine-structure transitions
can also become strong.
With appropriate observing strategies in combination with optical telescopes,
ALMA and/or SKA may be able to detect such lines,
offering a unique probe of physical conditions
in individual Pop III and early Pop II star forming regions.
We also remark on potential near-infrared absorption features due to electronic transitions of H$_2$.
\end{abstract}



\section{Introduction}
\label{sec:intro}

Gamma-ray bursts (GRBs) hold the promise of being powerful probes of the universe
at very high redshifts\footnote{In this paper, `very high redshift' refers to $z \ga 10$.}
(Loeb 2003, Djorgovski 2004).
Their presumably close link with the collapse of massive stars
indicates that they may occur
up to the earliest epochs of cosmic star formation (Bromm \& Loeb 2002, Schneider, Guetta \& Ferrara 2002),
and their extreme luminosities in broad wavebands
should allow them to be readily detected out to the highest relevant redshifts
(Lamb \& Reichart 2000, Ciardi \& Loeb 2000, Gou et al. 2004),
a realistic prospect with the launch of the SWIFT satellite mission (Gehrels et al. 2004).

The formation era of the first stars in the universe
is of considerable interest from many perspectives
(Barkana \& Loeb 2001, Bromm \& Larson 2004, Ciardi \& Ferrara 2005, Glover 2005 and references therein).
First, the pertinent physics in metal/dust-free conditions
should be markedly different from the present day situation,
most probably leading to predominant formation of very massive, Population III stars.
Second, such stars and their associated supernovae
are expected to induce crucial feedback effects on their surroundings,
including reionization and metal enrichment of the intergalactic medium,
which will then strongly affect subsequent star and galaxy formation.
After sufficient buildup of metals and dust,
a transition to the more familiar mode of Population II star formation should take place.
Finally, all these processes may be intimately related to
the early formation and evolution of supermassive black holes and quasars.

However, most of these inferences are theoretical
and little is known with certainty, as observational clues are very scarce.
Aside from indirect indications such as
evidence of early reionization in cosmic microwave background anisotropies (Spergel et al. 2003, 2006)
or possible contributions to the cosmic near-infrared background
(Santos, Bromm \& Kamionkowski 2002, Salvaterra \& Ferrara 2003; but see also Salvaterra \& Ferrara 2006), 
detailed observations of stellar populations in the corresponding redshift range $z \sim 10-30$
are still beyond the capability of current observational facilities.

One of the most direct observational signatures of the star formation process is
molecular line emission from the cooling, collapsing protostellar material,
which generally arise in the radio, submillimeter to infrared wavebands
and provide a powerful probe of star formation in the low redshift universe.
In fact, emission lines from CO have been observed from galaxies
with redshifts as high as $z \sim 6.4$ (Walter et al. 2003),
and next generation telescopes should certainly go deeper (Blain, Carilli \& Darling 2004).
However, it is uncertain whether this would be feasible much beyond $z \sim 10$,
since 1) in the standard cosmological picture of hierarchical structure formation,
the typical masses of collapsed, star-forming systems at the corresponding epochs
are expected to be much smaller, and
2) the metallicity, and hence the CO abundance, should be much lower
at earlier epochs of star formation.
Emission line fluxes from H$_2$ have also been estimated for very high-$z$ populations,
but they are typically so faint
as to be undetectable by all but the most sensitive future observational facilities
(e.g. Omukai \& Kitayama 2003, Mizusawa, Nishi \& Omukai 2005 and references therein;
see however Ciardi \& Ferrara 2001).

A different approach to probe cool, star-forming gas at high redshifts
is through absorption lines in the spectra of bright background sources such as quasars.
Up to now, molecular absorption lines have been successfully detected 
in only a handful of systems up to $z \sim 3.4$,
either through electronic transitions of H$_2$ in the rest-frame UV (e.g. Ledoux et al. 2003, Reimers et al. 2003),
or rotational transitions of molecules such as CO in the radio (e.g. Wiklind \& Combes 1999, Curran et al. 2004).
More and higher $z$ detections should undoubtedly follow
with future telescopes (Curran et al. 2004, Blain et al. 2004),
but the usefulness of quasars for such purposes may become limited at very high $z$:
as with galaxies, sufficiently bright quasars (those lying on the tail of the luminosity function)
may become much rarer in a standard, hierarchical cosmology,
if the close relation between the masses of black holes and their host galaxies
seen in the nearby universe also holds at early epochs.

For absorption line studies at very high $z$, GRBs may offer an important alternative to quasars.
Being the end result of stellar collapse events,
their typical luminosities are not expected to decrease strongly with redshift like quasars.
They could even be typically more luminous at high $z$ if black holes forming in GRB events
are more massive than at lower $z$ (M\'esz\'aros \& Rees 2003).
\footnote{However, whether Pop III stars can produce GRBs robustly is still an open question (Heger et al. 2003).}
GRBs also possess other advantages for this purpose,
such as the simpler, nonthermal intrinsic spectra devoid of strong emission lines,
and the weaker perturbative effects on their environments (Barkana \& Loeb 2003).
The main disadvantage is their transience; in order to obtain adequate absorption line spectra,
they must be observed while they are bright enough with suitable telescope settings,
which requires fast response to GRB alerts with appropriate preparations.
The implications of observing absorption in high-$z$ GRB spectra
due to intergalactic hydrogen or metals have been discussed for rest-frame optical/UV lines
(e.g. Miralda-Escud\'e 1998, Lamb \& Reichart 2000, Oh 2002, Furlanetto \& Loeb 2003,
Barkana \& Loeb 2003),
and radio 21-cm lines (Furlanetto \& Loeb 2002, Ioka \& M\'esz\'aros 2005).

After this paper was submitted, SWIFT discovered GRB050904 (Cusumano et al. 2006),
which was successfully followed up at various wavelengths
(e.g. Watson et al. 2006, Haislip et al. 2006, Bo\"er et al. 2006, Frail et al. 2006).
Spectroscopic observations by the Subaru telescope
have identified its redshift as $z=6.29$ (Kawai et al. 2006), the highest of any GRB to date,
and have already provided new constraints on cosmic reionization (Totani et al. 2006).
The event marks an important first step in the use of GRBs as probes of the high-$z$ universe,
and showcases the potential of future studies at even higher $z$.

This work focuses on the prospects for observing molecular and atomic absorption lines due to ambient protostellar gas
in the radio to IR spectra of high-$z$ GRBs,
and the implications thereof for probing star formation at very early epochs.
We first discuss in \S \ref{sec:aft} the broadband spectra of high-$z$ GRBs expected in standard afterglow models,
including both the forward and reverse shock components.
The details of the spectral model are described in \S \ref{sec:model}.
The continuum detectability in the radio to infrared bands is then assessed for
various observational facilities such as ALMA and SKA.
In \S \ref{sec:tau}, we present model calculations of primordial star-forming clouds
incorporating detailed chemical reaction networks,
from which the absorption line optical depths for different atomic and molecular transitions are evaluated.
\S \ref{sec:obs} is devoted to a discussion of the detectability of such absorption lines in GRB afterglow spectra
by telescopes including ALMA and SKA, and the consequent implications.
We conclude in \S \ref{sec:conc}.
The following cosmological parameters are adopted, consistent with {\it WMAP} results (Spergel et al. 2003):
$\Omega_m=0.27$, $\Omega_\Lambda=0.73$ and $h=0.71$.


\section{Broadband afterglow emission of high redshift GRBs}
\label{sec:aft}

\subsection{Model spectra and light curves}
\label{sec:splc}

The broadband afterglow emission accompanying GRBs can be described quite robustly 
as nonthermal emission from electrons accelerated in relativistic shocks,
which occur as the GRB-generating outflow is decelerated by the ambient medium
(for reviews, see van Paradijs, Kouveliotou \& Wijers 2000,
Piran 2004, Hurley, Sari \& Djorgovski 2004, Zhang \& M\'esz\'aros 2004, M\'esz\'aros 2006).
The nonthermal electrons generated at the forward shock (hereafter FS)
in the external medium give rise to synchrotron emission
which typically spans the radio to X-ray range
and decays relatively slowly,
being observable up to days or weeks after the burst for GRBs at typical redshifts ($z \sim 1$).
We describe the FS component as presented in the Appendix of Inoue (2004),
which generally follows standard afterglow model prescriptions
but also includes a treatment of synchrotron self-absorption at high ambient densities.
The values of the principal parameters for the FS are fiducially taken to be consistent with those obtained
from broadband modeling of well-observed afterglows (e.g. Panaitescu \& Kumar 2001, Yost et al. 2003):
isotropic-equivalent kinetic energy of blastwave $E=10^{53}$ erg,
ambient medium density $n=1 {\rm cm^{-3}}$,
half angle of jet collimation $\theta_j= 0.1$ rad,
fractions of post-shock energy imparted to relativistic electrons
$\epsilon_{e,f}=0.1$
and magnetic fields $\epsilon_{B,f}=0.01$, respectively,
and spectral index of electron distribution $p_f=2.2$.

Also important is the reverse shock (hereafter RS) that runs through the GRB ejecta,
which has a much shorter lifetime
but can involve amounts of energy dissipation comparable to the FS.
The associated synchrotron emission initially peaks at lower frequencies
and can be much brighter than the FS, typically dominating the optical and/or infrared bands for $z \sim 1$ GRBs.
Thereafter, it should rapidly fade and shift down in frequency toward the radio,
falling beneath the FS component a few days after the burst.
This RS component can provide a consistent explanation
of the optical flashes and radio flares observed in a number of GRBs
(e.g. Piran 2004, M\'esz\'aros 2006 and references therein).
Note that GRB050904 at $z=6.3$ also exhibited a bright optical flash
that can be interpreted as RS emission (Bo\"er et al. 2006).

Although the expected properties of the RS emission
have been discussed extensively with regard to the optical and/or radio bands,
very few works have explicitly addressed the details of the spectrum in between these bands
and the behavior of the spectral peak at the self-absorption frequency
(e.g. Nakar \& Piran 2004, McMahon, Kumar \& Piran 2006).
Therefore, here we employ our own formalism for evaluating the RS component,
as detailed in \S \ref{sec:model}.
As a unique feature,
synchrotron self-absorption is accounted for utilizing the appropriate absorption coefficient
instead of the commonly used blackbody limit approximation (e.g. Sari \& Piran 1999a,b, Kobayashi \& Zhang 2003),
which is likely to be less reliable due to ambiguities in the time-dependent area of the emitting surface.
This can be important for our purposes, as self-absorption generally determines the peak frequency and flux of the RS.
Here we only treat the time evolution after the reverse shock has crossed the GRB ejecta,
since the earlier evolution during shock crossing is not of immediate interest for the purposes of this work.
In addition to $E$ and $n$, further parameters important for the RS and their fiducial values are:
initial bulk Lorentz factor of GRB ejecta $\Gamma_0=100$,
intrinsic GRB duration $T=100$ s,
and the RS counterparts of $\epsilon_{e,f}$, $\epsilon_{B,f}$ and $p_f$,
respectively $\epsilon_{e,r}=0.1$, $\epsilon_{B,r}=0.01$ and $p_r=2.2$.
We will assume throughout that $\epsilon_{e,r}=\epsilon_{e,f}$ and $p_r=p_f$
and simply denote these as $\epsilon_e$ and $p$,
as this choice is physically plausible and consistent with
the current (limited) observations of optical flashes and radio flares.
In the case that we take $\epsilon_{B,r}=\epsilon_{B,f}$, we likewise simply write $\epsilon_B$.
The time-dependent, broadband afterglow spectra for our fiducial parameter burst at $z=1$ can be seen in Fig.\ref{fig:sz1}.

\begin{figure}
\centering
\epsfig{file=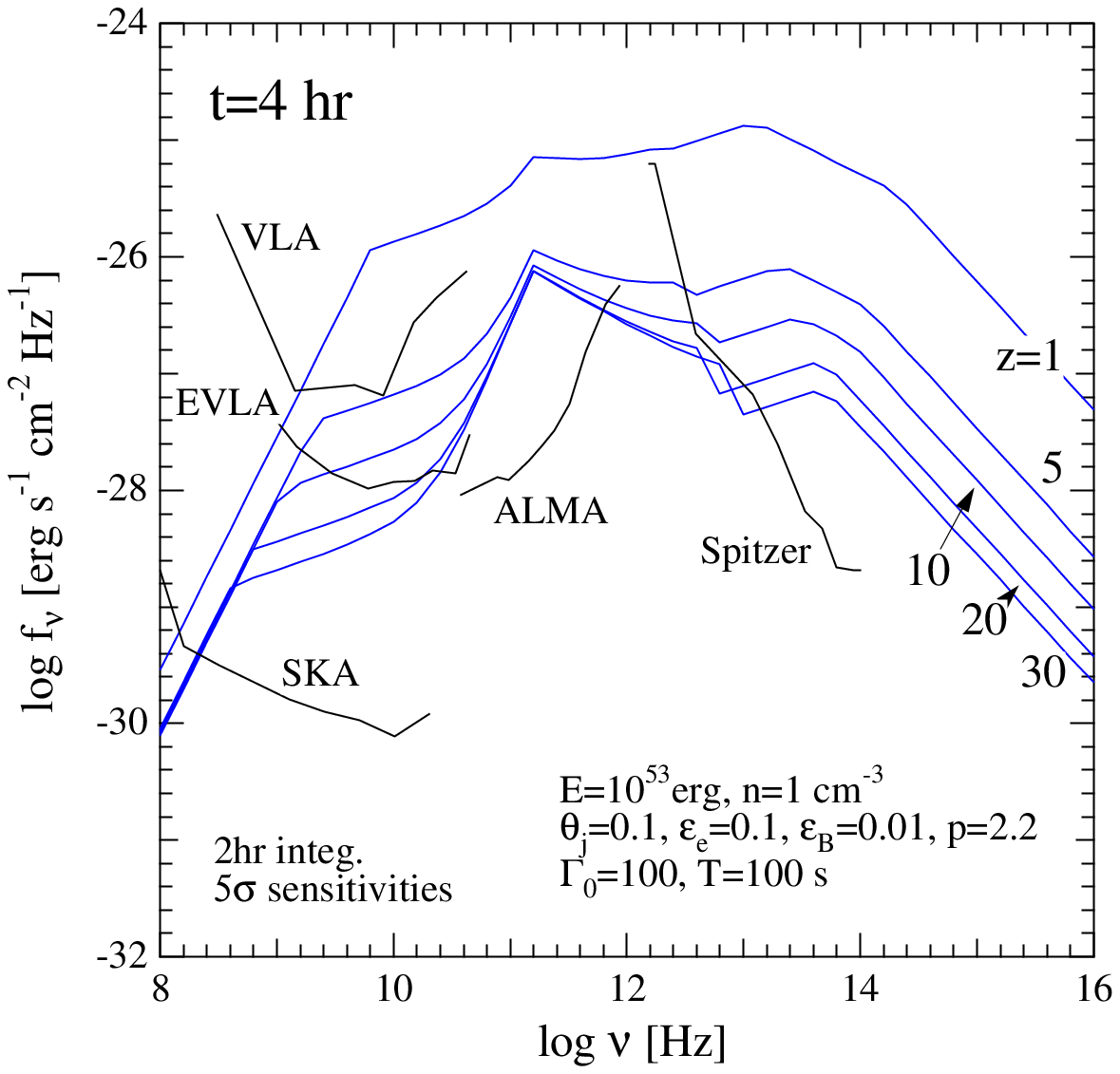,width=0.50\textwidth}
\epsfig{file=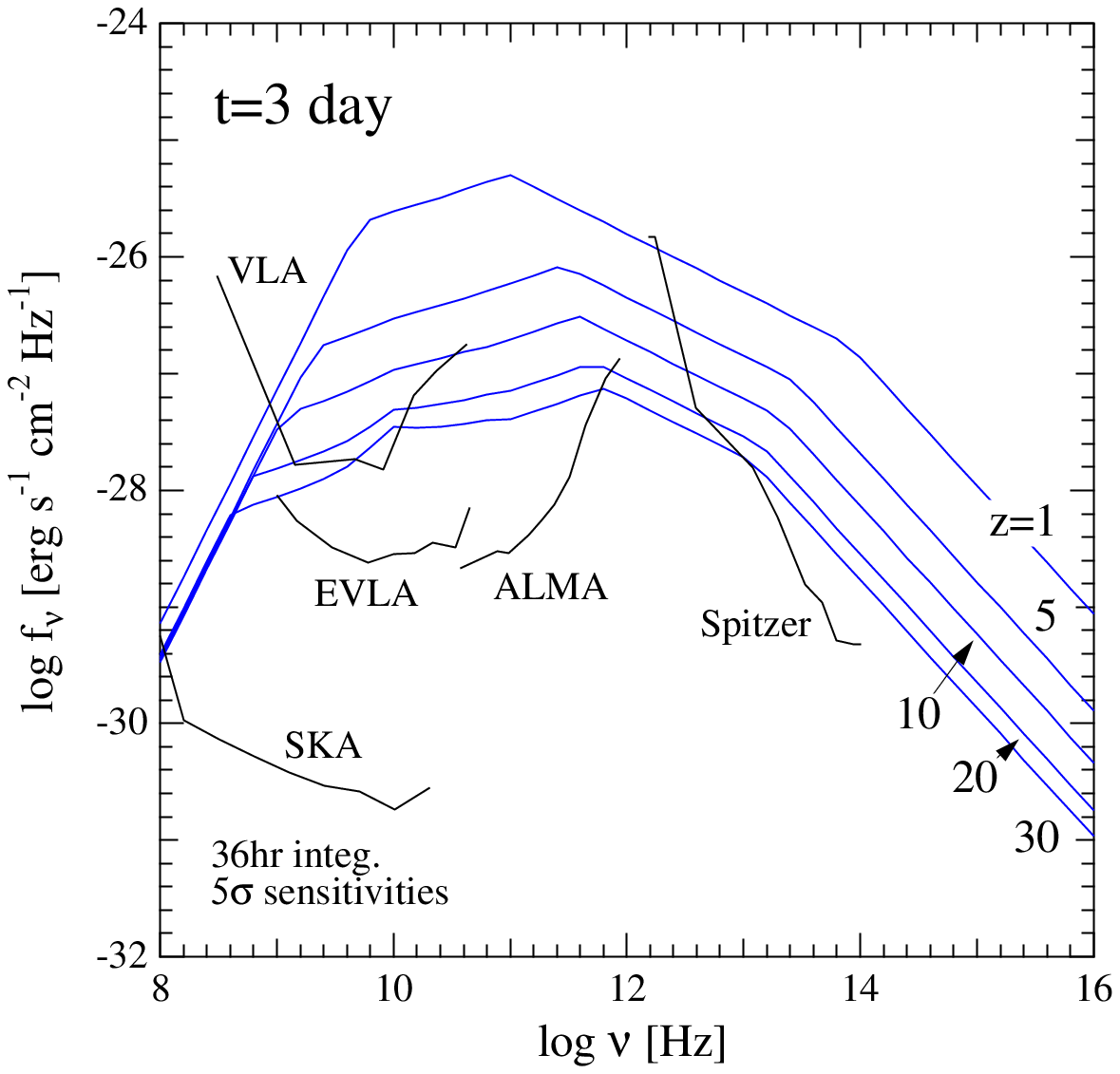,width=0.50\textwidth}
\caption{
(a) Broadband spectra of GRB afterglows at different $z$ as labelled
at fixed post-burst observer time $t=$ 4 hr, for our fiducial GRB.
Overlayed are 5 $\sigma$ continuum sensitivities of various observational facilities,
assuming integration times 50 \% of $t$.
(b) Same as a), but at $t=$ 3 days.
}
\label{fig:st4h3d}
\end{figure}

\begin{figure}
\centering
\epsfig{file=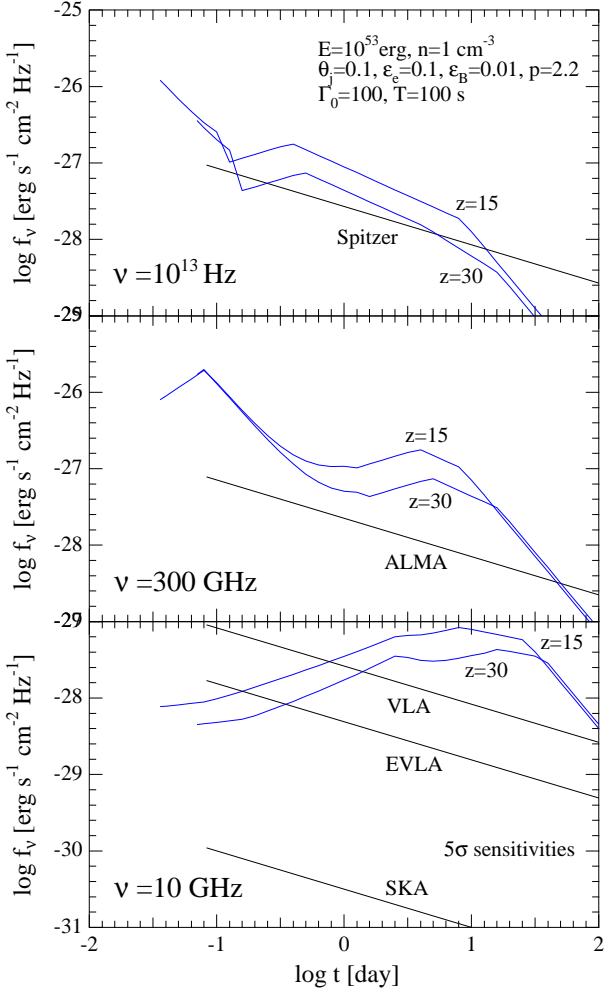,width=0.50\textwidth}
\caption{
Light curves of GRB afterglows at $z=15$ and $30$ at fixed observing frequencies
(a) $10^{13}$ Hz, (b) $300$ GHz and (c) $10$ GHz.
Only the post-RS crossing part ($t \ge t_{\times}$) is shown;
the flux at $t < t_\times$ is always less than at $t=t_\times$.
Overlayed are 5 $\sigma$ continuum sensitivities of observational facilities as labelled,
assuming integration times 50 \% of $t$.
}
\label{fig:lz1530}
\end{figure}

We thus evaluate the spectra and light curves of very high-$z$ GRBs.
Figs.\ref{fig:st4h3d} (a) and (b) show the broadband spectra 
of our fiducial GRB occurring at different redshifts from $z=1$ to 30,
seen at fixed post-burst observer times $t=$ 4 hours and 3 days, respectively.
Figs.\ref{fig:lz1530} (a)-(c) show the light curves at selected observing frequencies for GRBs at $z=$15 and 30.
As mentioned above, only the post-RS crossing portion ($t \ge t_\times$)
is displayed for the light curves; the fluxes at earlier $t$ should always be less than the first plotted point.

Within $t \sim$ 1 day, the two different components can be clearly distinguished in the spectra (see also Fig.\ref{fig:sz1}).
The FS contributes a broad, broken power-law spectrum extending from radio to X-ray frequencies,
with time-dependent break frequencies due to self-absorption, electron injection and cooling.
On the other hand, the RS spectrum spans a narrower range in the millimeter to far-infrared,
but can be significantly brighter in these bands, consisting of a self-absorption peak
and a cooling cutoff.
After $t \sim$ few days, the RS has mostly faded away and the FS dominates the whole spectrum.

Two striking features are evident for the RS component.
First, the peak frequency ($\nu \simeq$ 200 GHz at $t=$ 4 hr)
stays almost independent of $z$ despite the large cosmological redshifting.
Second, the flux at this frequency is also almost $z$-independent at $z \ga 5$
and remains remarkably bright, at milli-Jansky levels even from $z=30$.
In contrast, the FS component is progressively dimmer with $z$ at nearly all frequencies.
These properties follow from cosmological time dilation and the fast evolving nature of the RS emission;
a fixed observer time $t$ corresponds to an earlier rest-frame time at higher $z$,
when the RS component is much more prominent
and possesses a higher rest-frame peak frequency.
We can see this more specifically from the formulae in \S \ref{sec:model}
appropriate for our fiducial parameter set, which is
the thin shell, slow-cooling case with the break frequencies ordered as
$\nu_{m,r} < \nu_{a,r} < \nu_{c,r}$.
From Eqs.\ref{eq:taustn} and \ref{eq:tevtn}-\ref{eq:nua}, the RS peak frequency due to self-absorption depends on $t$ and $z$ as
$\nu_a \propto t^{\alpha_a} (1+z)^{-1-\alpha_a}$
with $\alpha_a \simeq -1.54 + 3.2/(p+4)$,
which is $\nu_a \propto t^{-1.03} (1+z)^{0.03}$ for $p=2.2$.
The dependence close to $t^{-1}$ for fixed $z$
means that the effects of time dilation almost exactly compensate for frequency redshift
when compared at fixed $t$, resulting in a near-constant observed peak frequency.
This behavior is quite insensitive to the value of $p$
(e.g. $\nu_a \propto t^{-1.05} (1+z)^{0.05}$ for $p=2.5$).
Similarly, from \ref{eq:fnu2}, the RS peak flux at $\nu_a$ is
$f(\nu_a) \propto t^{\alpha_f} (1+z)^{-1-\alpha_f} D_L^{-2}$
with $\alpha_f \simeq -0.97 + 1.6(p-1)/(p+4)$, where $D_L$ is the luminosity distance.
For $p=2.2$, this is $f(\nu_a) \propto t^{-1.3} (1+z)^{2.3} D_L^{-2}$.
The steep decay with $t$ at fixed $z$
implies a large flux increase with $z$ at fixed $t$,
counteracting the dimming with distance.
\footnote{Note that this trend continues only up to a certain $z$ above which the observer time $t < t_\times$,
when one starts seeing the pre-RS crossing part of the light curve and the flux decreases more strongly.}
The greatly enhanced significance of the RS emission for very high-$z$ GRBs
has also been discussed recently by Gou et al. (2004).
However, they did not account for self-absorption and ascertain the peak frequency of the RS,
as their main interest was in the near-infrared to X-ray domain.

Although to a lesser degree than the RS,
similar trends are exhibited by the FS emission as well,
as discussed by Lamb \& Reichart (2000) and Ciardi \& Loeb (2000).
For our fiducial burst, the FS is slow-cooling after $t \simeq$ 15 minutes,
in which case the FS peak frequency $\nu_p$
evolves with time as $\propto t^{-3/2}$ at fixed $z$ (e.g. Sari, Piran \& Narayan 1998).
This implies $\nu_p \propto (1+z)^{1/2}$ when compared at fixed $t$,
differing only by a factor $\simeq 2$ in the range $z=5-30$
(e.g. $\nu \simeq$ 250-570 GHz at $t=3$ days; Fig.\ref{fig:st4h3d} (b)).
The flux at fixed observing frequency and observer time is also of interest.
For $\nu \ge \nu_p$, $f_\nu$ falls with $t$ so that time-dilation leads to a higher flux at higher $z$ for fixed $t$;
on the other hand, the spectrum in this region falls with $\nu$,
so frequency redshift (K-correction) gives the opposite effect.
For $\nu \le \nu_p$, the converse is true because $f_\nu$ rises with $t$ and rises with $\nu$.
In either case, the effects of time-dilation and frequency redshift act against each other
to produce a flux which does not change greatly with $z$ at fixed $\nu$ and $t$.
This expectation has indeed been observationally confirmed by Frail et al. (2006) in the radio band, at least up to z=6.3.
Such effects should be even more dramatic for the RS due to its faster $t$-evolution.

\begin{figure}
\centering
\epsfig{file=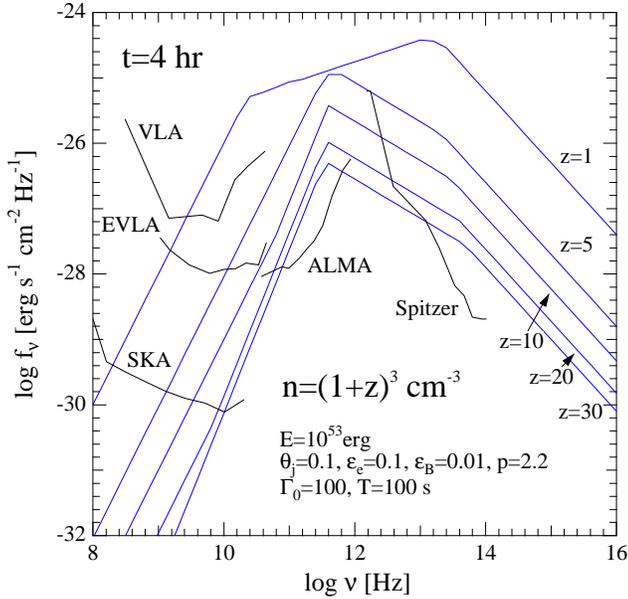,width=0.50\textwidth}
\caption{
Same as Fig.\ref{fig:st4h3d} (a), except that $n=(1+z)^3 {\rm cm^{-3}}$ has been assumed.
}
\label{fig:st4hnz3}
\end{figure}

We have studied comprehensively
how the fiducial results above are affected
when the main parameters are varied within a plausible range,
and the fact that the afterglow spectrum manifests a peak in the millimeter to infrared range
hours to days after the burst 
was found to be
a quite robust property.
As a demonstration of this statement,
we show in Fig.\ref{fig:st4hnz3} the results
when $n$ is assumed to depend on $z$ as $n=(1+z)^3 {\rm cm^{-3}}$,
which is one possible way that the interstellar medium density of high-$z$ galaxies may evolve
and a distinct alternative to our fiducial choice of  constant $n$
(see Inoue 2004 for more on this issue).
The rapid increase of $n$ with $z$ means that with the same $\epsilon_B$,
the post-shock magnetic fields (in this case for both FS and RS) grow larger.
This in turn implies that both synchrotron self-absorption and synchrotron cooling become stronger with $z$,
leading to sharply peaked spectra by suppressing the lower and higher frequency portions, respectively.
Intriguingly enough, when observed at fixed $t$,
the peak frequency is again almost independent of $z$ and lies in the submillimeter band at $t \sim$ few hours.

We mention that our spectral model of \S \ref{sec:model} is based on the approximation
of taking either the ultrarelativistic or Newtonian limit descriptions for the RS dynamics.
A more accurate evaluation may give quantitatively different results
in the trans-relativistic regime (e.g. Nakar \& Piran 2004).
However, our intention here is to provide illustrative model examples of broadband afterglow spectra
for their potential use in probing the high-$z$ universe,
rather than in comparisons with real observational data.
More detailed spectral modeling will thus be deferred to future work.

\subsection{Detectability of continuum emission}
\label{sec:detcont}

Noting the very high flux levels even at the highest $z$,
we assess the detectability of very high-$z$ afterglows
with current and future telescopes operating in the radio, submillimeter and infrared bands,
concentrating on the fiducial case.
In Figs.\ref{fig:st4h3d} - \ref{fig:st4hnz3}, the 5 $\sigma$ continuum sensitivities are overlayed
for the current Very Large Array (VLA) \footnote{http://www.vla.nrao.edu/},
the future Extended Very Large Array (EVLA) \footnote{http://www.aoc.nrao.edu/evla/},
Atacama Large Millimeter Array (ALMA) \footnote{//www.eso.org/projects/alma/; Wootten (2002)}, 
and Square Kilometer Array (SKA) \footnote{http://www.skatelescope.org; Carilli \& Rawlings 2004} facilities,
as well as the current Spitzer telescope \footnote{http://ssc.spitzer.caltech.edu/},
all assuming that the integration times are 50\% of $t$.

Particularly interesting is the millimeter to far-infrared range where the RS component falls.
Although observations here are generally more difficult than in neighboring bands
due to strong atmospheric absorption and other factors,
the prominent, mJy-level RS contribution allows the emission to be readily detectable out to $z \sim 30$
by ALMA with integration times of a few hours.
Even after the RS decays at $t \la$ 1 day, the light curves in these bands
generally brighten again at $t \sim$ several days, due to the passage of the FS peak frequency
(Figs. \ref{fig:st4h3d} (b) and \ref{fig:lz1530} (b)).
Although the flux of this second FS peak in the light curve is about an order of magnitude fainter
than the first RS peak,
the longer allowed integration times should make its detection by ALMA also feasible.
Note that the limited sensitivity of currently-operating telescopes in these wavebands
have so far enabled only the brightest, low-$z$ bursts to be detected
(Smith et al. 2005a, 2005b, Kuno et al. 2004, Kohno et al. 2005).

In the radio domain below few tens of GHz,
the RS is never really important and the flux is much less at early times.
However, the radio flux gradually brightens to 0.01-0.1 mJy
as the FS peak frequency reaches these frequencies at $t \ga$ 10 days (Fig.\ref{fig:lz1530}).
This should be well within reach of the generally superior sensitivities of radio facilities.
Even the current VLA, which has played a crucial role
in elucidating the properties of radio afterglows for GRBs observed to date (e.g. Frail 2003, Berger 2003),
may detect very high-z bursts as well, albeit with long integration times of a few days.
EVLA should be capable of doing this in several hours and SKA possibly in a few minutes.
These points were also discussed recently by Ioka \& M\'esz\'aros (2004).
In the mid- to near-infrared, Spitzer should also have a chance with $\sim $hour to $\sim $day integration times,
not to mention other ground based telescopes and the future James Webb Space Telescope (JWST)
at near-infrared wavelengths (Ciardi \& Loeb 2000, Gou et al. 2004).

The event rate of GRBs at high-$z$ is presently rather uncertain.
Extrapolating the redshift distribution of bursts observed by SWIFT so far,
Jakobsson et al. (2006) conclude that the fraction of bursts occurring at $z>5$ should be 7 \%-40 \%,
which is in agreement with recent theoretical expectations (e.g. Natarajan et al. 2005, Bromm \& Loeb 2006).
Bromm \& Loeb (2006) also predict that the rate at $z>10$ could be up to several per year.
Although a detailed assessment of the rate is beyond the scope of this paper,
it is conceivable that at least several detectable GRBs per year occur at $z \ga 5$, and some even beyond $z \sim 10$.
Their unambiguous identification, however, requires dedicated observational efforts (see Sec.4.2).

Successful detections of very high-$z$ GRB afterglows in the above bands would have important implications.
First, they may simply be the highest redshift sources of any kind to be detectable at these frequencies.
The very first stars in the universe are expected to appear at $z \ga 30$
(Miralda-Escud\'e 2003),
and these may end up in GRBs, providing the ``first observable light'' in the radio to far-infrared wavebands.
Second, the time-dilated evolution of very high-$z$ GRBs may
prove to be valuable for studying the physics of the GRB outflow itself,
by enabling the RS emission component to be observed more comprehensively than may be possible at low $z$.
Detecting the fast decaying RS signature for GRBs at typical distances ($z \sim 1$)
in the optical band has proved to be challenging,
with only a handful of successful cases (e.g. Akerlof et al. 2000, Roming et al. 2005).
While a number of explanations are possible for the paucity of detections (M\'esz\'aros 2006 and references therein),
a full understanding of the RS phenomenon may not be achieved without
covering the millimeter to infrared range, where a significant portion of the RS spectrum
might lie (Fig.\ref{fig:sz1} and \ref{fig:st4h3d}).
Observational facilities in these bands may possibly have a better chance of detecting the RS component
for higher $z$ GRBs, thanks to time dilation
which allows more time for telescope pointing.

However, the intriguing possibility which we highlight in the following
is that some GRB afterglows may be bright enough in the radio to infrared bands
for high resolution spectroscopic observations 
of molecular and atomic absorption lines due to ambient cold gas,
which could have far-reaching implications for exploring
the earliest epochs of star formation in the universe.


\section{Primordial star-forming clouds and absorption line optical depths}
\label{sec:tau}

We now turn to discuss how much absorption may be expected by molecular and atomic fine-structure lines
in star-forming environments at very high redshifts, within which GRBs may arise.
Our estimates are based on Omukai (2000, 2001) and Omukai et al. (2005), who constructed models
of cooling, gravitationally collapsing clouds with varying levels of metallicity and external irradiation.
Although the dynamics was treated with a one-zone approximation,
the thermal evolution and non-equilibrium chemistry were calculated in detail
by following a large number of reactions
among the atomic and molecular species of H, He, C and O as well as dust grains.
The calculations here follow Omukai et al. (2005),
who has improved upon Omukai (2000, 2001) by including
the chemistry of deuterium, in particular that of the crucial molecule HD,
as well as the effects of the cosmic microwave background (CMB).
The potential influence of the radiation from the GRB itself is not included here,
but is discussed in \S \ref{sec:molenv}.

A dynamically collapsing, spherical cloud without rotation or magnetic fields
should possess a central core region of roughly constant density $\rho$
and size given by the local Jeans length $\lambda_J=\pi c_s/(G \rho)^{1/2}$,
where $c_s$ is the sound speed (e.g. Omukai \& Nishi 1998).
Focusing on this core region, the gas collapse is assumed to proceed
on the free fall timescale $t_{ff}=(3\pi/32G \rho)^{1/2}$
(note that the gravity due to dark matter is important only in the earliest stages of the collapse).
The initial number density and temperature are chosen to be $n=0.1 {\rm cm^{-3}}$ and $T=300$ K in all cases;
however, the calculated subsequent evolution for a given composition
quickly becomes independent of these values, as discussed in Omukai (2000).
Defining the concentration $y_x=n_x/n_H$ of a particular species $x$ as
the ratio of its number density $n_x$ to that of hydrogen nuclei $n_H$,
the abundances of He and D nuclei are fixed to their primordial values
$y_{\rm He}=9.72 \times 10^{-2}$ and $y_{\rm D}=4 \times 10^{-5}$.
The metallicity $Z$ is parameterized by its fraction with respect to the local interstellar value $Z_{\sun}$,
assuming that the relative abundances of heavy elements do not change.
The model of Pollack et al. (1994) is adopted for the dust grain properties,
and the dust-to-gas ratio is assumed to be proportional to $Z$.
The redshift of the protostellar cloud is fiducially taken to be $z=15$;
this determines the CMB temperature, although its effect is minor
for clouds of zero or very low metallicity. 

Another important parameter is the initial value of the H$_2$ concentration $y_{\rm H2,0}$
before cloud collapse.
For the very first generation of stars forming in H$_2$ cooling-dominated mini-halos with virial temperatures $T_v<10^4$K,
the pre-collapse H$_2$ concentration should be close to the primordial intergalactic value
$y_{\rm H2,0} \simeq 10^{-6}$,
since gas falling onto mini-halos is only weakly shocked.
However, for later generation stars forming in more massive, atomic cooling-dominated halos with $T_v>10^4$K,
the initial H$_2$ concentration can be as high as $y_{\rm H2,0} \simeq 10^{-3}$.
Here the gas should first be strongly shocked and ionized at the virialization shock,
which then lead to effective H$_2$ formation mediated by abundant free electrons
left over from non-equilibrium recombination in the post-shock layer
(Shapiro \& Kang 1987, Oh \& Haiman 2002).
Similar considerations apply to gas that has been shocked and compressed by supernova blastwaves
from a preceding generation of stars (Ferrara 1998, Bromm, Yoshida \& Hernquist 2003).
We consider below the two cases of $y_{\rm H2,0}=10^{-6}$ and $10^{-3}$.

The optical depth to absorption for each molecular or atomic line transition 
from level $i$ to $j$ is evaluated by
\begin{eqnarray}
\tau_{ij} = {A_{ij} c^3 \over 8\pi \nu_{ij}^3} [{g_i \over g_j}n_j - n_i] {l_{sh} \over v_D} ,
\end{eqnarray}
where $A_{ij}$ is the Einstein coefficient for spontaneous transitions,
$\nu_{ij}$ is the transition frequency, $g_i$ is the statistical weight of level $i$,
$v_D=(2k_B T/\mu_x m_H)^{1/2}$ is the velocity dispersion for a species with molecular weight $\mu_x$,
and $l_{sh}=\min(R_c, \Delta s_{th})$ is the shielding length,
taken to be the smaller of either the core radius $R_c=\lambda_J/2$
or the Sobolev length $\Delta s_{th}=v_D/(dv/dr)=3 v_D t_{ff}$.
This gives an estimate for the line optical depth through the core region of the cloud,
the precise value depending on the exact geometry and specific line-of-sight.

\subsection{Zero metallicity clouds}
\label{sec:tauzero}

\begin{figure}
\centering
\epsfig{file=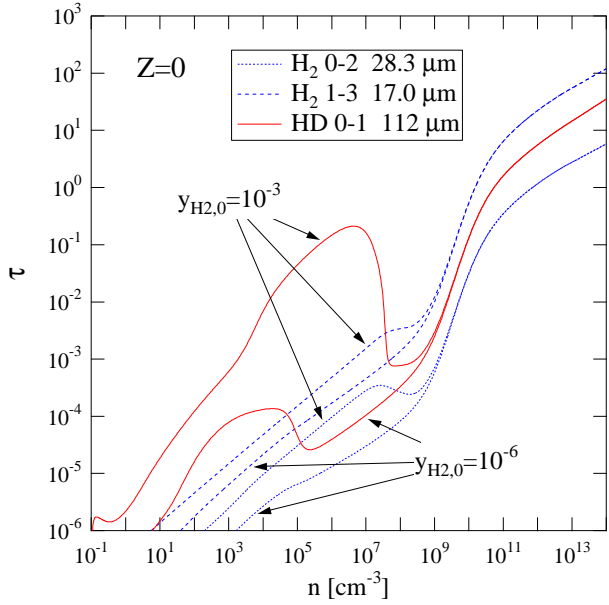,width=0.50\textwidth}
\caption{
Absorption line optical depth $\tau$ versus number density $n$
for the rotational transitions $J=0 \rightarrow 2$ and $J=1 \rightarrow 3$ of H$_2$
and $J=0 \rightarrow 1$ of HD,
in protostellar clouds with $z=15$, metallicity $Z=0$
and initial H$_2$ concentrations $y_{\rm H2,0}=10^{-6}$ and $10^{-3}$.
}
\label{fig:Z0tauH}
\end{figure}

We first consider a completely metal-free ($Z=0$) cloud.
Fig.\ref{fig:Z0tauH} shows the optical depth $\tau$
for the two lowest-lying rotational transitions of H$_2$
($J=0 \rightarrow 2$, $\nu=$ 28.3 $\mu$m and $J=1 \rightarrow 3$, $\nu=$ 17.0 $\mu$m)
and the lowest-lying rotational transition of HD
($J=0 \rightarrow 1$, $\nu=$ 112 $\mu$m)
as a function of the total number density $n$ in a $Z=0$ cloud at redshift $z=15$.
Two different initial H$_2$ concentrations have been assumed,
$y_{\rm H2,0}=10^{-6}$ and $10^{-3}$, as discussed above.
There is a one-to-one relation between the time after collapse and the density (and hence the temperature);
a given value of $n$ corresponds to a particular time during the cloud evolution
when the core has a size $\lambda_J \simeq 1.1 \times 10^{18} (T_2/n_6)^{1/2} {\rm cm}$
and is collapsing on a timescale $t_{ff} \simeq 5.2 \times 10^4 n_6^{-1/2} {\rm yr}$,
where the notations $n=10^6 n_6 {\rm cm^{-3}}$ and $T=100 T_2 K$ have been used.
The evolutionary trajectories in the $n-T$ plane for selected cases can be found in Omukai et al. (2005).

We see that the optical depths for H$_2$ in all situations
are generally quite low unless very high densities are reached.
That for HD is similarly low if $y_{\rm H2,0}=10^{-6}$.
Thus, for the very first generation protostellar clouds in mini-halos,
strong absorption lines are not expected
unless the line of sight passes through the small, innermost region of the collapsing core.

Contrastingly, in the case of $y_{\rm H2,0}=10^{-3}$,
the HD optical depth manifests an interesting behavior,
increasing rapidly with $n$ and approaching $\tau \sim 1$
at relatively low densities $n \simeq 10^6 - 10^7 {\rm cm^{-3}}$,
and then declining temporarily at higher $n$.
Initially, the temperature keeps decreasing and becomes low enough ($T \la 150$K)
for a large fraction of the available D to form HD,
whereas the H molecular fraction still remains at $\simeq 10^{-3}$.
HD also has a much larger $A$ coefficient thanks to its dipole moment, which H$_2$ lacks.
These two facts allow the HD optical depth to exceed that for H$_2$ by a few orders of magnitude.
However, HD reaches local thermodynamic equilibrium above $n \simeq 10^5 {\rm cm^{-3}}$,
and thereafter the temperature keeps rising
until much of the HD becomes dissociated, leading to the drop in $\tau$.
Note that around this temperature minimum, cooling is dominated by HD
and allows the gas to reach lower temperatures ($T \sim 50 K$) than is possible with H$_2$ alone.

Thus, for later generation star formation in relatively massive halos
or supernova shells where $y_{\rm H2}=10^{-3}$ can be realized,
HD may cause sufficiently strong absorption even in metal-free environments,
potentially offering a unique probe of Population III star formation (\S \ref{sec:impli}).

\subsection{Low metallicity clouds}
\label{sec:taulow}

\begin{figure}
\centering
\epsfig{file=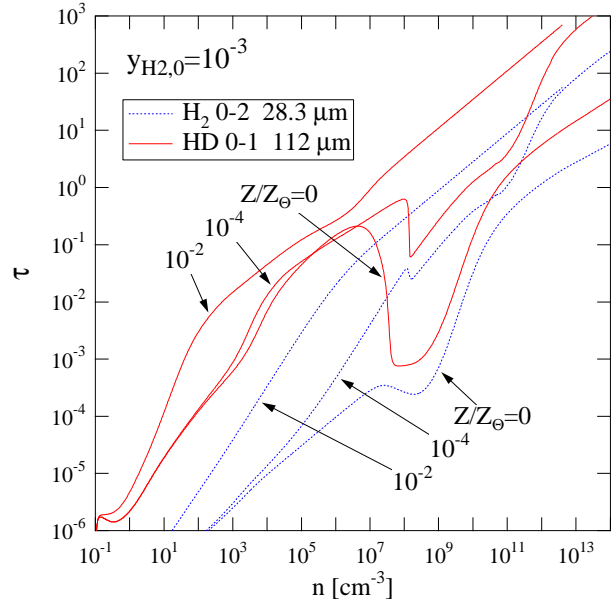,width=0.50\textwidth}
\caption{
Absorption line optical depth $\tau$ versus number density $n$
for the rotational transitions $J=0 \rightarrow 2$ of H$_2$ and $J=0 \rightarrow 1$ of HD,
in protostellar clouds with $z=15$, $y_{\rm H2,0}=10^{-3}$ and metallicities $Z/Z_{\sun}=0$, $10^{-4}$ and $10^{-2}$.
}
\label{fig:Z42tauH}
\end{figure}

\begin{figure}
\centering
\epsfig{file=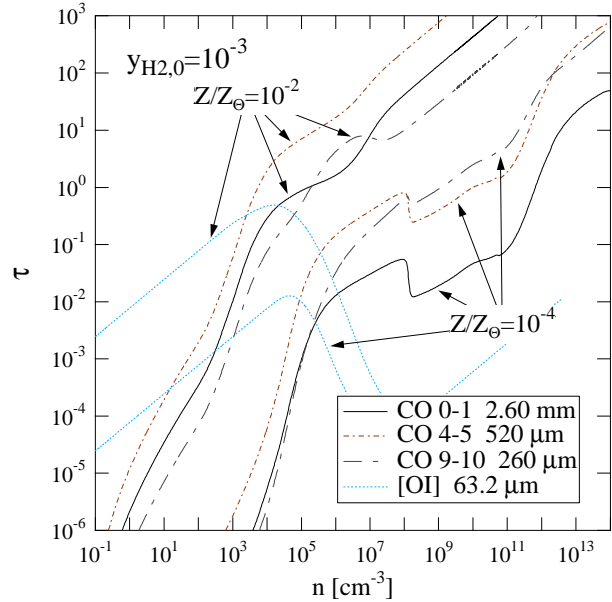,width=0.50\textwidth}
\caption{
Absorption line optical depth $\tau$ as a function of number density $n$
for the rotational transitions $J=0 \rightarrow 1$,
$J=4 \rightarrow 5$ and $J=9 \rightarrow 10$ of CO,
and the
lowest-lying fine structure transition of [OI],
in protostellar clouds with $z=15$, $y_{\rm H2,0}=10^{-3}$ and $Z/Z_{\sun}=10^{-4}$ and $10^{-2}$.
}
\label{fig:tauCO}
\end{figure}

The line optical depths for clouds with $y_{\rm H2,0}=10^{-3}$
and different levels of metallicity are shown in Figs.\ref{fig:Z42tauH} and \ref{fig:tauCO}.
Higher $Z$ allows more efficient H$_2$ formation through dust grain surface reactions
and consequent cooling to lower temperatures.
This also induces HD formation, which cools the gas even further
by becoming the dominant coolant for some intermediate range of densities depending on $Z$ (Omukai et al. 2005).
Unlike the $Z=0$ case, HD is not immediately destroyed at $n \ga 10^5 {\rm cm^{-3}}$
because the temperature can remain low.
All these facts contribute to a generally higher $\tau$ with higher $Z$
for both H$_2$ and HD, as can be seen in Fig.\ref{fig:Z42tauH}.

Fig.\ref{fig:tauCO} displays the optical depths
for different rotational transitions of the CO molecule
($J=0 \rightarrow 1$, $\nu=$ 2.6 mm; $J=4 \rightarrow 5$, $\nu=$ 520 $\mu$m and
$J=9 \rightarrow 10$, $\nu=$ 260 $\mu$m).
When $Z=10^{-2}Z_{\sun}$, the optical depths of CO lines can exceed unity at densities
as low as $n \simeq 10^3 - 10^5 {\rm cm^{-3}}$.
Even for a very low metallicity of $Z=10^{-4}Z_{\sun}$,
the higher-excitation transitions of CO can become strong at $n \ga 10^7 {\rm cm^{-3}}$.
Since the critical metallicity for transition from a Pop III, high mass dominated mode of star formation
to a Pop II, low mass dominated one is expected to be in the range $Z_{crit}/Z_{\sun}=10^{-6}-10^{-4}$
(Bromm et al. 2001, Schneider et al. 2002, 2003, Omukai et al. 2005),
CO lines should constitute a valuable probe of the very early Pop II era,
and possibly of the crucial transition epoch as well.
We also see that for a given $Z$, the relative strengths of different rotational transitions 
vary with $n$ (and hence $T$) in interesting ways.
If more than one CO line can be measured for the same cloud,
the line ratios should provide a useful diagnostic of the cloud physical properties (\S \ref{sec:impli}).

Also shown in Fig.\ref{fig:tauCO} is the optical depth to
the lowest-lying atomic fine structure transition of [OI] ($\nu=$ 63.2 $\mu$m),
which approaches $\sim 1$ at $n \simeq 10^4 {\rm cm^{-3}}$ if $Z=10^{-2} Z_{\sun}$.
That for [CII] is about an order of magnitude less.
These absorption lines should be important for probing the earliest stages of the collapse,
as [OI] and [CII] are thought to be the main cooling agents at this time
(\S \ref{sec:impli}).

We mention that the one-zone models presented here are intended only
to provide order of magnitude estimates of the absorption line optical depths
in primordial protostellar clouds.
More realistic evaluations should be carried out on the basis of 3-D simulations
(e.g. Abel, Bryan \& Norman 2002, Bromm, Coppi \& Larson 2002,
Yoshida et al. 2006, Gao et al. 2007),
if the chemistry of HD or metals/dust can be fully incorporated within them in the future.


\section{Observability of absorption lines and implications}
\label{sec:obs}

In \S \ref{sec:aft}, we saw through standard relativistic blastwave models
that very high-$z$ GRB afterglows should have spectra robustly peaking around the submillimeter band,
initially at a few hours after the burst with milli-Jansky flux levels due to the reverse shock,
and then for a second time several days afterwards with somewhat less flux due to the forward shock.
In \S \ref{sec:tau}, model estimates were presented for absorption line optical depths
in primordial protostellar clouds which may host very high-$z$ GRBs,
demonstrating potentially appreciable absorption due to HD even in metal-free gas,
as well as strong absorption from CO and [OI] in the case of sufficiently metal-enriched gas.
Bringing together these discussions,
here we assess the observability of molecular and atomic absorption lines
due to cool, protostellar gas in very high-$z$ GRB afterglows and its implications.

\subsection{Molecular gas in GRB environments}
\label{sec:molenv}

The first important issue to be addressed is how much molecular gas
may actually exist in the vicinity of GRBs when they occur.
GRBs likely result from core collapse at the endpoint of a massive star's life,
and these progenitor stars should be born embedded
in the dense centers of star-forming clouds like those discussed in \S \ref{sec:tau}.
However, a massive star emits copious amounts of UV radiation during its lifetime,
which is very effective in photodissociating molecules within the star's vicinity, before the GRB goes off.
In fact, radiation from a single massive star may be enough to destroy
the entire molecular content of a host mini-halo with $T_v<10^4$K,
at least from simple considerations assuming spherical symmetry
(Omukai \& Nishi 1999, Glover \& Brand 2001; see also Johnson, Greif \& Bromm 2007 for recent 3-D simulations).
Radiation from the GRB itself can also contribute to photodissociation;
however, the total number of UV photons is at most $\sim 10^{61}$ even for a bright UV flash (e.g. Draine \& Hao 2002),
which is less than that expected from the progenitor star (e.g. Ciardi et al. 2001).
Gamma-rays and X-rays are even less important because of
the fewer number of photons and smaller absorption cross sections.
At any rate, for mini-halos with $T_v<10^4$K,
the line optical depths should be small anyway
because of the low $y_{\rm H2,0}$ (\S \ref{sec:tauzero}),
so it may be unlikely that any absorption lines would be observable 
for GRBs emerging from the very first generation of stars in the universe.

The prospects may be better for GRBs from later generation stars forming in halos with $T_v>10^4$K.
First, larger absorption line depths are realized due to the higher $y_{\rm H2,0}$ (\S \ref{sec:tauzero}),
as well as the larger gas masses in such halos.
Second, the gas in these large halos can collapse into numerous clouds which then fragment into multiple clumps, 
each of which can become a star-forming core (e.g. Oh \& Haiman 2002, Bromm, Coppi \& Larson 2002).
Such situations are reminiscent of massive star formation
in giant molecular clouds in the Galaxy today (e.g. Lada \& Lada 2003).
Within a clump hosting a massive star,
destruction of molecules is likely to be effective due to photodissociation
as well as heating by photoionization and/or stellar winds.
However, other clumps within the same cloud that are sufficiently far away
and have yet to form a star inside should be free from such adverse effects,
because the molecular column densities are large enough for these clumps
to be self-shielded from external UV radiation.
For the example of our zero-metallicity cloud with $y_{\rm H2,0}=10^{-3}$,
regions that approach $n \sim 10^6 {\rm cm^{-3}}$ and $\tau \sim 1$ for the HD line
correspond to column densities $N_{\rm HD} \sim 10^{18}  {\rm cm^{-2}}$,
which far exceed the minimum column density required for self-shielding,
$N_{\rm HD} \sim 10^{13}  {\rm cm^{-2}}$ (Draine \& Bertoldi 1996).
Thus, such pristine clumps lying in the foreground of a GRB can potentially cause
interesting absorption lines of HD.
Note, however, that the radius $\sim 0.3$ pc and lifetime $\sim 5 \times 10^4$ yr
expected for the high $\tau$ core (\S \ref{sec:tauzero})
may be much smaller compared to those for the whole clump,
estimated to be $\sim 10$ pc and $2 \times 10^6$ yr
for an initial clump density $n \sim 10^3 {\rm cm^{-3}}$
as appropriate in halos with $T_v>10^4$K (Oh \& Haiman 2002).
The probability of these clump cores occupying the GRB line of sight
depends on their uncertain spatial and temporal distribution and is presently difficult to predict,
but quantitative and realistic estimates should be tractable in the future through 3-D simulations.

A further possibility arises if the photodissociation region (PDR) around a massive star
can sweep up the surrounding material and trigger further star formation.
Such situations have indeed been observed inside our Galaxy (Deharveng et al. 2003, Hosokawa \& Inutsuka 2005),
and may also be possible in very high-$z$ environments, although this is yet to be fully investigated.
Newly collapsing clumps on the PDR periphery would have suffered little photodissociation,
and a relatively high $y_{\rm H2,0}$ is plausible due to the shock compression.
The chances may be better for such clumps to induce HD absorption lines.

After metal enrichment has proceeded to above a critical value $Z_{crit}/Z_{\sun}=10^{-6}-10^{-4}$,
cooling by metals and dust becomes important.
Besides the higher collapse densities attained through more efficient cooling,
the buildup of dust also
leads to more effective formation of H$_2$ on dust grain surfaces,
as well as greater shielding against external UV radiation.
For example, in our case with $Z/Z_{\sun}=10^{-2}$,
regions with $n \ga 10^4 {\rm cm^{-3}}$ and $\tau \ga 1$ for the lower-excitation CO lines
have H$_2$ column densities $N_{\rm H2} \sim 10^{21}  {\rm cm^{-2}}$,
which is sufficient for shielding of CO, not to mention H$_2$ and HD (Lee et al. 1996).
The expected optical depths can be high through most of the collapsing cloud,
especially for CO lines,
those for HD are also higher compared to when $Z=0$ (\S \ref{sec:taulow}).
It is therefore quite probable that strong molecular absorption lines 
are formed in such metal-enriched environments.

While having no destructive capability,
the radio to infrared emission from the GRB
can potentially alter the level populations of ambient molecules through radiative excitation.
Its importance can be approximately gauged through the simple example of a two-level system,
described in detail in Appendix B of Stahler \& Palla (2004).
We assume a homogeneous medium of density $n$
composed of particles with two levels separated by energy $h \nu_0$,
immersed in an isotropic radiation field of intensity $I$ in steady state conditions.
Radiative excitation plays a significant role when $n \la n'_{cr}  \equiv n_{cr} (1+c^2 I/2h \nu_0^3)$,
where $n_{cr} \equiv A_0/\gamma_0$ is the critical density
above which collisions dominate and the system attains LTE in the absence of an external radiation field,
being expressible in terms of the Einstein coefficient for spontaneous emission $A_0$
and the coefficient for collisional deexcitation $\gamma_0$ between the two levels.
In all cases of our interest, $n \gg n_{cr}$.
We may define a critical radiation intensity above which radiative excitation is important,
\begin{eqnarray}
I_{cr} \equiv {2 h \nu_0^3 \over c^2} {n \over n_{cr}},
\end{eqnarray}
to be compared with the mean intensity $I_{GRB}$ of our fiducial afterglow
at a distance $d$ from the GRB in the molecular cloud.

For the conditions relevant to the $J=0 \rightarrow 1$ HD line ($\nu_0$=2.68 THz)
in our $Z=0$, $y_{\rm H2,0}=10^{-3}$ case (\S \ref{sec:tauzero}),  
$n/n_{cr} \sim 30$ and $I_{cr} \sim 8.5 \times 10^{-9} {\rm erg \ s^{-1} cm^{-2} Hz^{-1} sr^{-1}}$.
When the afterglow is brightest at this rest-frame frequency at $t \sim 10$ min (Fig.\ref{fig:sz1}),
$I_{GRB} \sim 10^{-9} {\rm erg \ s^{-1} cm^{-2} Hz^{-1} sr^{-1}} (d/10 {\rm pc})^{-2}$,
which drops thereafter by 1 to 2 orders of magnitude hours to days after the burst.
Therefore, the afterglow radiation will have no bearing on the excitation states of HD unless they lie at $d \ll 10$ pc.
As discussed above, the vicinity of the GRB should be devoid of molecules anyway due to the actions of the progenitor star.
The effect could be more relevant for some of the CO lines in metal-enriched clouds (\S \ref{sec:taulow}).
For example, where $\tau \ga 1$ for the $J=0 \rightarrow 1$ CO line ($\nu_0$=115 GHz)
in the $Z/Z_{\sun}=10^{-2}$ case,
$n/n_{cr} \sim 100$ and $I_{cr} \sim 2.2 \times 10^{-12} {\rm erg \ s^{-1} cm^{-2} Hz^{-1} sr^{-1}}$.
At the corresponding frequency, the fiducial afterglow has
$I_{GRB} \sim 10^{-10} {\rm erg \ s^{-1} cm^{-2} Hz^{-1} sr^{-1}} (d/10 {\rm pc})^{-2}$
hours to days after the burst,
so CO molecules within $d \sim 100$ pc may have their lower excitation states affected.
A realistic evaluation of such effects is outside the scope of this work and set aside for future studies,
requiring spatially-dependent calculations with time-varying, anisotropic radiation fields.
We proceed below under the assumption that the molecules are located at $d \ga 10-100$ pc
where radiation excitation is negligible,
in line with the above discussion of multiple foreground clumps in large clouds and/or clumps precipitated by PDR shocks.

\subsection{Detectability of absorption lines}
\label{sec:detabs}

Given that the circumstances envisioned above are realized for at least some GRBs,
the detectability of the absorption lines with different observational facilities can be appraised.
The absorption line widths should be dominated by velocity gradients
from the bulk infall motion in the collapsing cloud,
which is roughly
\begin{eqnarray}
\Delta v \sim \lambda_J/t_{ff} \simeq 6.7 \ {\rm km/s} \ T_2^{1/2} .
\end{eqnarray}
Since the temperatures are generally $T \simeq 30-100$ K when large $\tau$ values occur, 
a velocity resolution of at least 3 km/s would be necessary to detect such lines.
We note that diffuse regions in which turbulent motions are important
are unlikely to have sufficient optical depth for absorption lines.

\begin{figure}
\centering
\epsfig{file=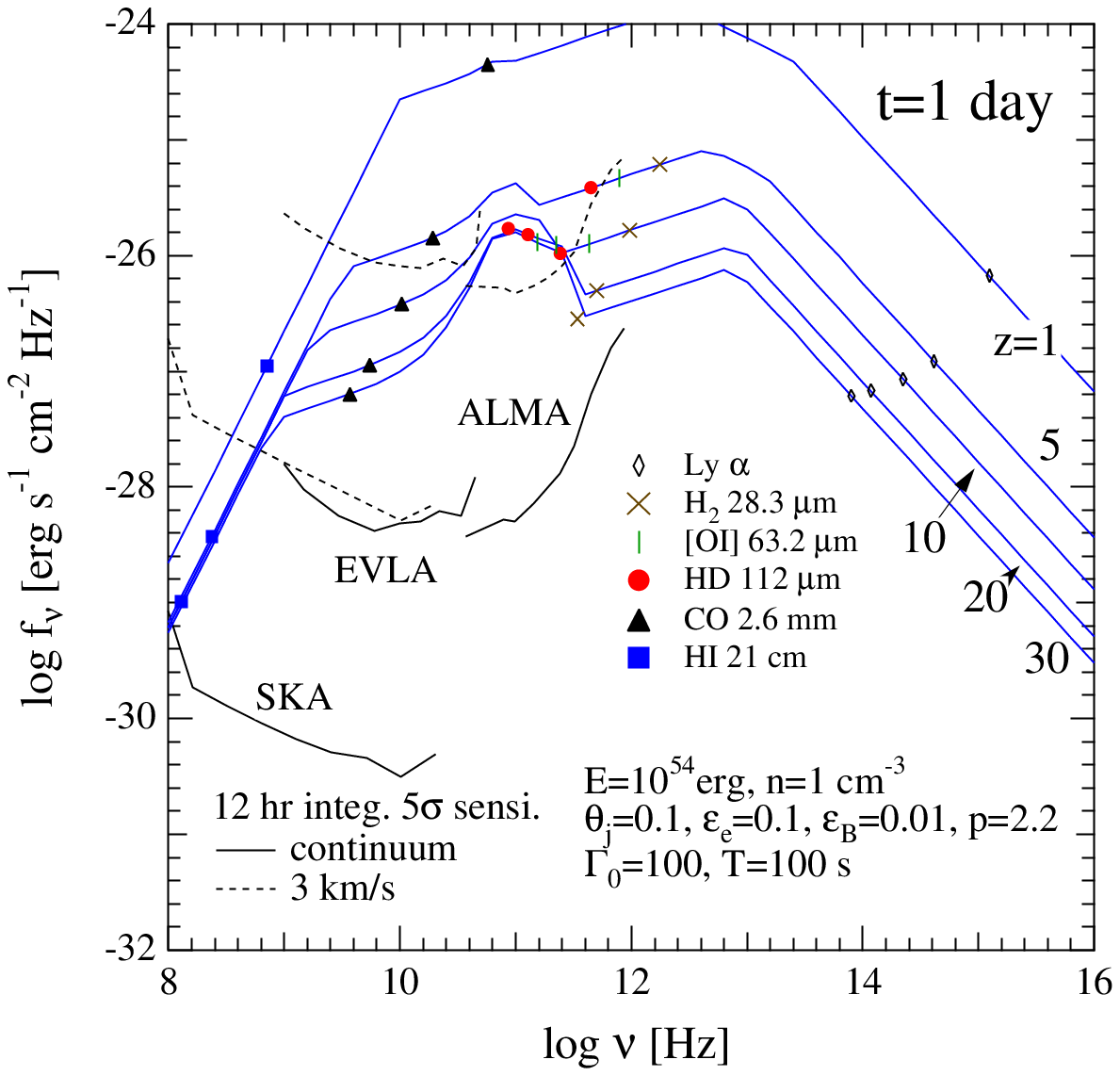,width=0.50\textwidth}
\epsfig{file=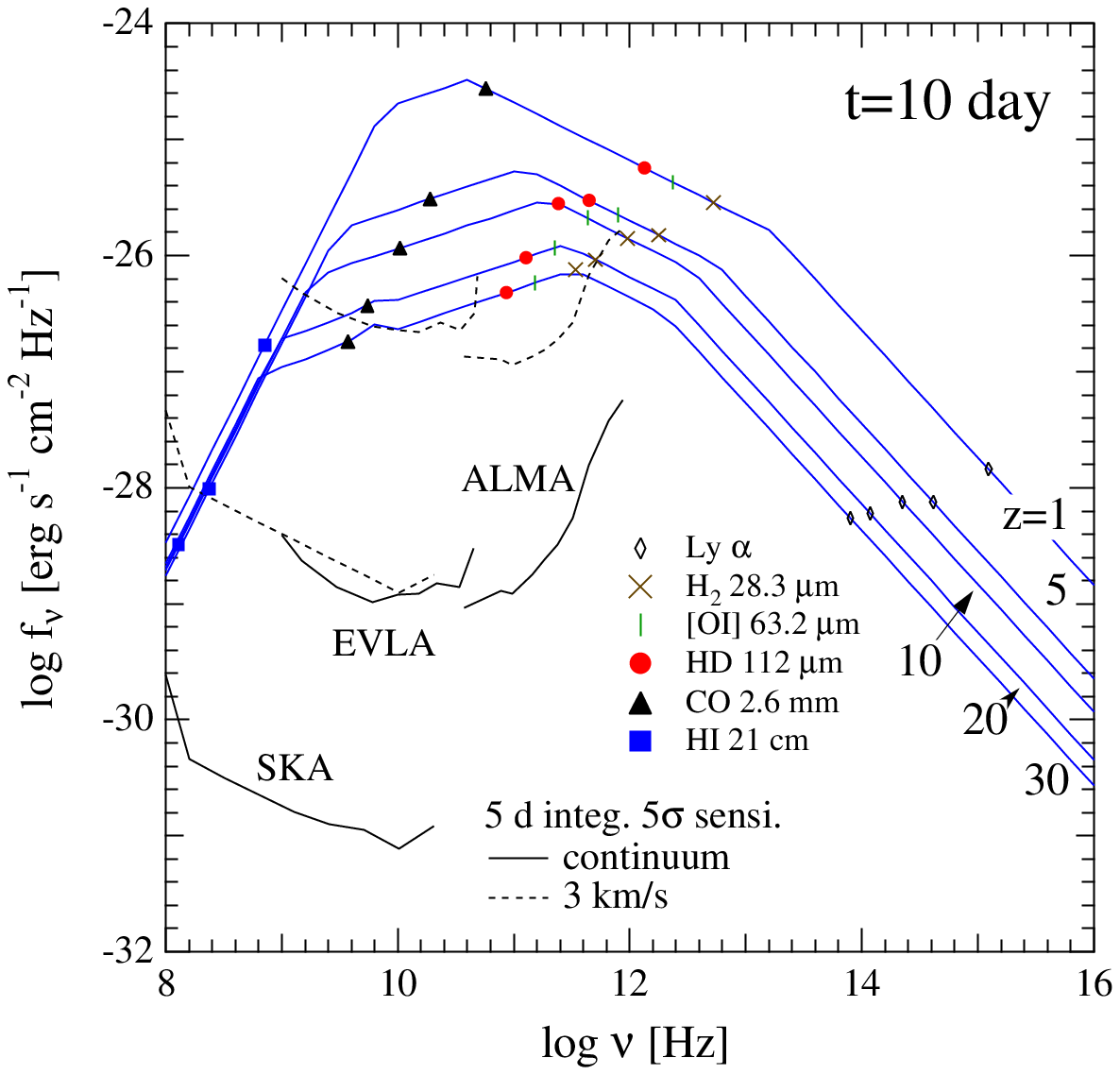,width=0.50\textwidth}
\caption{
Broadband spectra of GRB afterglows with fiducial parameters except for $E=10^{54}$ erg,
at different $z$ as labelled and fixed post-burst observer time (a) $t=$ 1 day and (b) 10 days.
The redshifted frequencies for the lowest-lying transitions of 
H$_2$ (crosses), HD (circles), CO (triangles) and [OI] (vertical bars),
as well as the Ly $\alpha$ (diamonds) and HI 21cm (squares) transitions are indicated on each spectra.
Overlayed are 5 $\sigma$ continuum and 3 km/s resolution spectroscopic sensitivities
of various observational facilities, assuming integration times 50 \% of $t$.
}
\label{fig:tt1d10d}
\end{figure}

\begin{figure}
\centering
\epsfig{file=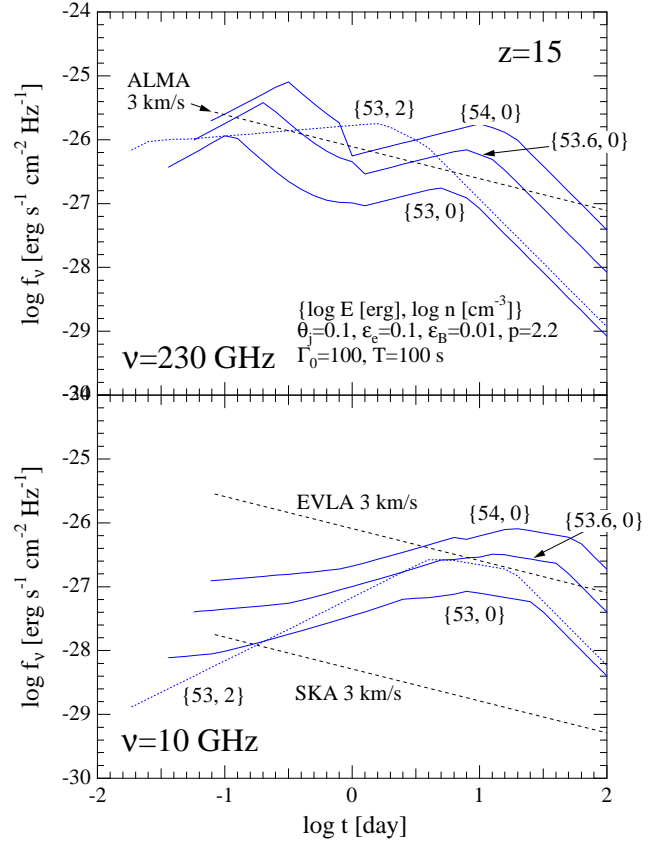,width=0.50\textwidth}
\caption{
Light curves of GRB afterglows with different $E$ and $n$ (as labelled by their logarithmic values),
at $z=15$ and at fixed observing frequencies
(a) $230$ GHz and (b) $10$ GHz (only the post-RS crossing part $t \ge t_{\times}$ is shown).
Overlayed are 5 $\sigma$, 3 km/s resolution spectroscopic sensitivities of observational facilities as labelled,
assuming integration times 50 \% of $t$.
}
\label{fig:lz15en}
\end{figure}

In Fig.\ref{fig:tt1d10d},
we show the broadband spectra of a relatively bright burst with $E=10^{54}$ erg (all other parameters being fiducial),
at different redshifts from $z=$1 to 30 and post-burst observer times $t=$ 1 and 10 days.
Marked on each spectra are the redshifted frequencies for the lowest-lying transitions of H$_2$, HD, CO and [OI].
To be compared are the 5-$\sigma$ sensitivities of ALMA, EVLA and SKA,
for both continuum and spectroscopic observations at 3 km/s resolution,
assuming integration times 50 \% of $t$.
Note that this spectroscopic sensitivity is for optically thick absorption lines;
to detect lines with small or moderate optical depths at the same resolution and significance,
the required sensitivity (or alternatively, the required background continuum flux) is higher
by a factor $(1-e^{-\tau})^{-1}$ ($\sim \tau^{-1}$ for $\tau \ll 1$).
Fig.\ref{fig:lz15en} displays the light curves of GRBs at $z=$15 with different values of $E$ and $n$,
at observing frequencies $\nu$=10 GHz (near the low-excitation CO transitions)
and 230 GHz (near the HD or [OI] transitions),
in comparison with the spectroscopic sensitivities of the aforementioned telescopes.

The redshifted absorption lines for the lower-excitation transitions of CO are mostly in the radio regime,
while those for HD and [OI], as well as the higher-excitation transitions of CO
all fall in the submillimeter band.
There are interesting prospects for all of these lines to be detectable.
For GRBs with sufficiently large $E \ga 4 \times 10^{53}$ erg,
ALMA may in fact have two opportunities to detect the submillimeter lines,
first due to the passage of the bright RS peak at $t \la$ 1 day,
and second due to the passage of the FS peak around $t \sim$ 10 days
(however, see below on practical difficulties with the first RS peak).
Note that large isotropic equivalent energies $E \la 10^{54}$ erg
are not uncommon in observed GRBs at high redshifts (Amati 2004),
including GRB050904 at $z=6.3$ (Cusumano et al. 2006).
\footnote{The true total energy may be less due to collimation,
but collimation effects in the afterglow should not be important in very high-$z$ afterglows until $t \ga$ 10 days.}
A burst with average $E$ but relatively high $n \ga 10^2 {\rm cm^{-3}}$
should also manifest a bright enough FS component at $t \sim$ 1 day
(Fig.\ref{fig:lz15en}).

In the radio, the chances are even better.
The CO lines should be detectable by EVLA
for GRBs with $E \ga 4 \times 10^{53}$ erg at $t \ga$ 10 days.
SKA, with its outstanding sensitivity, should do the same for typical energy bursts in less than a day.
With longer integration times, SKA may even be capable of measuring weak lines with $\tau \la 0.1$,
as well as strong lines in subluminous bursts.

Besides adequate sensitivity and spectral resolution,
some important technical constraints pertaining to radio/submm observations
must be spelled out here.
One is that interferometers such as ALMA or SKA
require sufficiently bright calibrator sources
near the intended targets in the sky in order to carry out absolute flux measurements.
Depending on the actual distribution of suitable calibrators,
which will only be known after ALMA begins operation,
some restrictions may be imposed on regions of the sky in which GRBs can be observed.

However, the most important caveat regards frequency coverage.
For radio and submillimeter observations,
achieving high frequency resolution as is necessary here
mandates that the instantaneous bandwidth be made correspondingly narrow.
For example for ALMA, the number of spectral channels
that the correlator can provide for an analyzing bandwidth $B$ is
$128 (B/{\rm 2 GHz})^{-1}$ per baseline for two polarizations.
A velocity resolution of 3 km/s at observing frequency $\nu$=230 GHz
implies that the maximum allowed bandwidth would be $4B \simeq 3.2$ GHz
(T. Wilson, private communication),
so at a given time, only a small portion of the spectrum can be covered.
This is not a great concern when searching for lines in steady sources,
as a broad spectral range can be scanned over time.
However, transient sources such as GRBs pose a problem:
even if one aims for a particular line with a definite rest-frame frequency,
its redshift must be known beforehand to good accuracy (for the above example, to within 1 \%)
in order to begin observing at the appropriate frequency while the source is still bright.

Thus a crucial prerequisite is that good redshift information for the GRB is quickly available by other means.
Spectrographs on moderate-sized optical telescopes are currently indeed able
to measure redshifts for GRB afterglows through metal absorption lines
with accuracies better than fractions of a percent within a day or so after the burst,
at least up to $z \sim 6$ (e.g. Fynbo et al. 2006 and references therein).
Doing so for bursts at higher $z$ is complicated by contamination from infrared airglow,
but can be achieved through spectroscopy of the Ly break by large telescopes with near-infrared capabilities.
For example, spectrographs similar to the CISCO instrument on the Subaru telescope
can obtain a $J$-band spectrum (covering the Ly break at $z=10$) at resolving power R=200
with S/N=10 in 1 hour for a 3 $\mu$Jy source (T. Yamada, private communication),
which is more than sufficient for the afterglow fluxes discussed here (Fig.\ref{fig:tt1d10d}).

Rapid dissemination of such redshift information over the GRB Coordinate Network
is already being carried out extensively.
In the ALMA era, successor missions to SWIFT
such as EXIST \footnote{http://exist.gsfc.nasa.gov/} or
ECLAIRs \footnote{http://www.oamp.fr/ECLAIRS/} should be in operation.
In synergy with gamma-ray and optical/infrared telescopes,
successful detections of radio/submm absorpton lines in very high-$z$ GRB afterglows should be feasible.
For this purpose, the FS shock peak at $t \ga$ several days should be much more suitable
than the early decaying RS shock peak, because
1) more time is allowed before the redshift is measured and the appropriate radio/submm observations can begin,
and 2) the FS emission should be more robust compared to the RS emission,
which can be weak for some bursts with strong RS magnetic fields.

In addition to the above lines,
we have also indicated the 21 cm hyper-fine transition frequency (1.4 GHz)
in the spectra of Fig.\ref{fig:tt1d10d}.
Although strong 21 cm absorption may potentially be caused by 
host galactic disks at high-$z$  (Furlanetto \& Loeb 2003),
it can be readily seen that even with the sensitivity of SKA and a bright GRB,
detecting such absorption lines in GRB afterglows should be very difficult unless $z \la 10$.
This is because the redshifted 21 cm line lies
deep in the low frequency, self-absorbed part of the afterglow spectrum
where the flux is at micro-Jansky levels. 
These inferences were mentioned in Inoue (2004) and also agree with Ioka \& M\'esz\'aros (2004).
For studying 21 cm absorption at very high-$z$,
radio-loud quasars may thus be more promising sources
(Carilli, Gnedin \& Owen 2002, Haiman, Quataert \& Bower 2004),
although the existence and number of such objects at $z \ga 10$ is quite uncertain at this moment.

\subsection{Implications of detecting absorption lines}
\label{sec:impli}

For each kind of absorption line discussed above,
there are tremendous implications if they can actually be detected.

As mentioned in \S \ref{sec:tauzero},
HD can play a crucial role in the cooling of zero or low metallicity gas in addition to H$_2$.
In some cases it can actually become the dominant coolant, cooling the gas down to few 10s of K
and possibly even allowing the formation of relatively low mass stars
(Uehara \& Inutsuka 2000, Nakamura \& Umemura 2002, Flower 2002,
Nagakura \& Omukai 2005, Johnson \& Bromm 2006).
The true importance of HD cooling in primordial star formation is not yet clear ,
but observations of HD absorption lines in high-$z$ GRBs by ALMA should provide a quantitative test.
If it can be demonstrated at the same time
that the metal content of the gas is so low as to be insignificant for cooling
(either through metal lines in the near-infrared or CO lines),
it will offer a very unique signature of Population III star formation
(albeit not for the very first generation of stars in mini-halos,
but for later generation stars in more massive halos or triggered by supernovae/PDR shocks).

It was shown that some high-excitation CO lines (e.g. $J=4 \rightarrow 5$ to $J=9 \rightarrow 10$)
can be relatively strong even when $Z \ga 10^{-4} Z_{\sun}$ (\S \ref{sec:taulow}),
which may be near the critical metallicity for transition in star formation from
high mass dominated, Population III to low mass dominated, Population II
(Bromm et al. 2001, Schneider et al. 2002, 2003, Omukai et al. 2005).
This points to such lines (potentially observable by ALMA) being an important probe of the earliest phases of Pop II,
and perhaps even of the critical transition era,
which is of great interest from many viewpoints
(e.g. Mackey, Bromm \& Hernquist 2003, Scannapieco, Schneider \& Ferrara 2003, Salvaterra \& Ferrara 2003).

At higher metallicity levels, the low-excitation CO lines (e.g. $J=0 \rightarrow 1$ to $J=4 \rightarrow 5$)
can be very strong,
being readily observable by SKA and in some cases by EVLA.
In fact, SKA should be sensitive enough to detect several different rotational transitions of CO
from the same cloud during the time that the GRB afterglow is bright (Fig.\ref{fig:lz15en} (b)).
Measurement of line ratios will give information on physical conditions
such as temperature and density for individual clouds, which may be impossible to do in any other way.
Note that detecting the HD lines may also be possible for metal-enriched clouds
and provide important additional information.

During the earliest stages of collapse, [OI] and [CII] are the main cooling agents
for metal-enriched clouds (Omukai 2000, Bromm \& Loeb 2003, Omukai et al. 2005).
Observing [OI] in absorption by ALMA will therefore constitute an interesting diagnostic of this phase,
and may also offer a different perspective from CO lines on the Pop III to Pop II transition epoch.

\subsection{Near-infrared absorption by H$_2$}
\label{sec:nir}

We briefly touch upon the possibility of observing electronic absorption features by H$_2$ in the near-infrared.
Absorption due to electronic transitions in the rest-frame UV are potentially very interesting
because the $A$ coefficients are much larger and the required column densities much smaller 
compared to ro-vibrational transitions.
Indeed, as mentioned in \S \ref{sec:intro},
electronic absorption features have already been used for damped Lyman alpha systems at $z \la 3.4$
as an important probe of H$_2$, sensitive down to column densities $N_{\rm H2} \simeq 10^{14} {\rm cm^{-2}}$
(Ledoux et al. 2003, Reimers et al. 2003),
and in one case even of HD (Varshalovich et al. 2001).
However, absorption from cold H$_2$ in the Lyman-Werner bands (912-1110 \AA)
will not be useful above $z \sim 6$,
becoming completely obscured  by Gunn-Peterson troughs shortward of Ly $\alpha$ from intergalactic HI (White et al. 2003).

On the other hand, H$_2$ in the local environment of GRBs may be more interesting.
We have mentioned in Sec.\ref{sec:molenv} that the bulk of the molecular gas in sufficiently distant clumps
should be self-shielded from photodissociation by external UV radiation.
However, the layers closest to the GRB can be affected by UV photons from the burst.
Draine (2000) and Draine \& Hao (2002) have shown that
if H$_2$ with $N_{H2} \ga 10^{18} {\rm cm^{-2}}$ lies
at distances $R \la 10^{19} {\rm cm}$ around a GRB,
which can indeed be the case for our clouds discussed above,
the optical-UV emission from the GRB itself can pump the ambient H$_2$ into vibrationally excited states.
This will in turn induce strong, time-dependent absorption in the GRB spectrum up to rest-frame 1705 \AA,
with characteristic features that can be identified through spectroscopy at resolution $R \ga 300$.
This may then provide an observable signature of H$_2$ even at $z \ga 6$
appearing in the near-infrared longward of Ly $\alpha$, which we have also marked in Fig.\ref{fig:tt1d10d}.
In fact, such a signature may have already been seen in GRB050904 at $z=6.3$ (Haislip et al. 2006).
Although a detailed discussion is beyond the scope of this paper,
such electronic absorption features may offer interesting prospects
for telescopes with fast response and near-infrared spectroscopic capabilities
as a unique and valuable probe of H$_2$ at very high-$z$.


\section{Conclusions}
\label{sec:conc}

The salient points of this work are summarized.
We have evaluated the time-dependent, broadband afterglow spectra of very high-$z$ GRBs
using standard relativistic blastwave models with both forward and reverse shock components,
paying particular attention to the latter.
It was found that for a plausible range of physical parameters and external medium densities,
GRBs in the redshift range $z \sim 5 - 30$ exhibit spectra
which robustly peak in the millimeter to infrared range with milli-Jansky flux levels
when observed a few hours after the burst, mainly due to the reverse shock.
Following the decay of the reverse shock component, the light curve in these bands can brighten again
for several days afterwards due to the forward shock.
The continuum emission should be readily observable up to $z \sim 30$ and beyond
by telescopes such as ALMA, EVLA, SKA and other facilities in the radio, submillimeter and infrared.

For relatively bright GRBs, high resolution spectroscopic measurements
of atomic and molecular absorption lines due to ambient protostellar gas may be possible.
We have utilized models of zero- and low-metallicity star-forming clouds to show that
under certain conditions, sufficiently strong absorption may be caused by transitions of HD, CO and [OI].
Such atoms and molecules should be immune to dissociation or excitation
by radiation from the progenitor or the GRB as long as they lie 10-100 pc away from the GRB site.
HD lines may constitute a unique probe of metal-free, Pop III star formation,
albeit not for the very first generation of stars in mini-halos,
but for later generation stars forming in more massive halos
or on the peripheries of photodissociation regions.
CO and [OI] lines can offer valuable information on the early Pop II era,
possibly including the Pop III to Pop II transition epoch,
and also provide diagnostics of physical conditions in the protostellar gas through line ratios.
A cooperative observing strategy with gamma-ray and optical telescopes should enable
the detection of HD, [OI] and high-excitation CO lines by ALMA,
and low-excitation CO lines by SKA (and in some cases, EVLA). 
Absorption features due to H$_2$ in the near-infrared may also be a potentially interesting signature.

Because of their transient nature,
the detection of radio/submm absorption lines in GRB afterglows may be challenging in practice.
However, they offer us a unique view into star-forming regions at very high redshifts ($z \ga 10$),
which may be quite difficult through emission lines or absorption lines in quasars
for the reasons discussed in \S \ref{sec:intro}.
On the other hand, at lower redshifts observationally accessible by emission lines or quasar absorption lines,
their steady fluxes give them an advantage.
Lines in absorption and emission also provide different kinds of information,
the former on local conditions along specific lines of sight,
and the latter on globally integrated properties such as total mass.
In these respects, different approaches using emission lines, absorption lines in quasars and in GRBs
should all be valuable in complementary ways for investigating early star formation.

\section*{Acknowledgments}
We acknowledge valuable discussions regarding the technical aspects of observations
with Tom Wilson, Richard Tuffs, Jochen Greiner,
Seiichi Sakamoto, Ryohei Kawabe, Kazushi Sakamoto and Toru Yamada.
Dale Frail, Peter M\'esz\'aros, Takashi Hosokawa and Shuichiro Inutsuka are thanked for helpful comments.
This work was partially supported by the auspices of the European Research and Training Network
``Gamma-Ray Bursts: An Enigma and A Tool''.


\appendix

\section{Broadband afterglow emission}
\label{sec:model}

The forward shock (FS) component of the GRB afterglow emission is modeled as in the Appendix of Inoue (2004).
This generally follows standard prescriptions for the synchrotron emission from electrons
accelerated by a relativistic shock decelerating self-similarly in the ambient medium
(e.g. Sari, Piran \& Narayan 1998, Panaitescu \& Kumar 2000),
and includes a treatment of synchrotron self-absorption at high ambient densities.
As explained in \S \ref{sec:aft}, the main parameters are
$E$, $n$, $\theta_j$, $\epsilon_{e,f}$, $\epsilon_{B,f}$ and $p_f$.
The spectrum consists of power-law segments adjoining break frequencies at
$\nu_{m,f}$ and $\nu_{c,f}$, corresponding respectively
to the injection energy $\gamma_{m,f}$ and cooling energy $\gamma_{c,f}$ in the electron distribution,
as well as $\nu_{a,f}$, the self-absorption frequency.
The peak flux $f_{p,f}$ occurs at $\nu_{p,f}=\min(\nu_{m,f},\nu_{c,f})$ when $\nu_{a,f}<\nu_{p,f}$,
and at $\nu_{a,f}$ reduced by self-absorption when $\nu_{a,f}>\nu_{p,f}$.
The break frequencies and the peak flux evolve with time in characteristic ways
through the evolution of the shock radius, bulk Lorentz factor and post-shock magnetic field.
For more details and specific expressions, the reader is referred to Inoue (2004).

For the reverse shock (RS),
additional parameters are $\Gamma_0$, $T$, $\epsilon_{e,r}$, $\epsilon_{B,r}$ and $p_r$,
of which we take $\epsilon_{e,r}=\epsilon_{e,f} \equiv \epsilon_e$ and $p_r=p_f \equiv p$ (\S \ref{sec:aft}).
The dynamical behavior of the RS can be approximately classified into two cases,
depending on the relation between the observed GRB duration $T(1+z)$
which characterizes the initial ejecta width $\Delta=cT$,
and
\begin{eqnarray}
t_\Gamma&=&\left(3E \over 4\pi n m_p c^2 \Gamma_0^2 \right)^{1/3} {1+z \over 2 \Gamma_0^2 c}\\
        &=& 195 \ {\rm s} \ E_{53}^{1/3} n^{-1/3} \Gamma_{0,2}^{-8/3} (1+z) ,
\label{eq:tgam}
\end{eqnarray}
the observed timescale for the forward shock to sweep up $1/\Gamma$ of the ejecta mass from the ambient medium
(Sari \& Piran 1995, Kobayashi 2000).
In the numerical expression,
we have used the notation $Q=10^p Q_p$ to represent physical quantity Q.
When $T(1+z)>t_\Gamma$, referred to as the thick shell case,
the reverse shock becomes relativistic in the frame of the unshocked ejecta
before the RS crosses the whole ejecta at $t_\times \simeq T(1+z)$,
and the ejecta is decelerated to bulk Lorentz factor $\Gamma_\times \simeq \Gamma_c$, where
\begin{eqnarray}
\Gamma_c&=&\left(3E \over 32\pi n m_p c^2 \Delta^3 \right)^{1/8}\\
        &=& 128 E_{53}^{1/8} n^{-1/8} T_{2}^{-3/8}.
\label{eq:gamc}
\end{eqnarray}
When $T(1+z)<t_\Gamma$, referred to as the thin shell case,
the reverse shock becomes marginally relativistic in the frame of the unshocked ejecta
just when the RS finishes crossing the ejecta at $t_\times \simeq t_\Gamma$,
and the ejecta is not significantly decelerated, $\Gamma_\times \simeq \Gamma_0$.

As with the FS, the RS spectrum is a broken power-law characterized
by its injection frequency $\nu_{m,r}$ and cooling frequency $\nu_{c,r}$
corresponding to break energies $\gamma_{m,r}$ and $\gamma_{c,r}$ in the electron distribution,
along with the self-absorption frequency $\nu_{a,r}$ and peak flux $f_{p,r}$.
Its shape and evolution can be modeled in an approximate way
by relating $\nu_{m,r}$, $\nu_{c,r}$ and $f_{p,r}$
with their FS counterparts at the shock crossing time $t=t_{\times}$
(Kobayashi \& Zhang 2003, Zhang, Kobayashi \& M\'esz\'aros 2003).
From equality of the bulk Lorentz factor and energy density in the post-shock regions across the contact discontinuity,
we can derive the relations
\begin{eqnarray}
\nu_{m,r}(t_\times) \simeq \nu_{m,f}(t_\times) {\Gamma_0^2 \over \Gamma_\times^4} {\cal R}_B^{1/2}
\label{eq:numrel}
\end{eqnarray}
and
\begin{eqnarray}
\nu_{c,r}(t_\times) \simeq \nu_{c,f}(t_\times) \left(1+Y_f \over 1+Y_r\right)^2 {\cal R}_B^{-3/2} ,
\label{eq:nucrel}
\end{eqnarray}
where ${\cal R}_B=\epsilon_{B,r}/\epsilon_{B,f}$,
and $Y_f$ and $Y_r$ are the Compton parameters for the FS and RS, respectively.
As was assumed for the FS in Inoue (2004),
$Y_r$ is approximated by its fast cooling value $Y_r=(\sqrt{4\epsilon_{e,r}/\epsilon_{B,r}+1}-1)/2$
at all times.
Additionally, from energy conservation between the kinetic energies in the initial ejecta and the shocked ambient medium,
one gets the relation
\begin{eqnarray}
f_{p,r}(t_\times) \simeq f_{p,f}(t_\times) {\Gamma_\times^2 \over \Gamma_0} {\cal R}_B^{1/2} .
\label{eq:fprel}
\end{eqnarray}

Hereafter, all quantities pertain to the RS, and we drop the subscript $r$ for brevity.
After shock crossing ($t>t_\times$), the time evolution of the above quantities can be obtained
through the evolution of the shock radius $R$, bulk Lorentz factor $\Gamma$, number density ${\cal N}$,
energy density ${\cal E}$ and magnetic field $B$ of the shocked ejecta as given in Kobayashi (2000).
In the thick shell case,
\begin{eqnarray}
R \propto t^{1/8}, \Gamma \propto t^{-7/16}, {\cal N} \propto t^{-13/16}, {\cal E} \propto t^{-13/12}, B \propto t^{-13/24}.
\label{eq:gevtk}
\end{eqnarray}
Since $\gamma_m \propto {\cal E}/{\cal N}$, $\nu_m \propto \Gamma B \gamma_m^2$,
and $\nu_c$ evolves in the same adiabatic way as $\nu_m$ after shock crossing (Sari \& Piran 1999a,b, Kobayashi 2000),
\begin{eqnarray}
\nu_m, \ \nu_c \propto t^{-73/48} \simeq t^{-1.52} .
\label{eq:nuevtk}
\end{eqnarray}
Similarly, as $f_p \propto \Gamma B {\cal N} R^3$,
\begin{eqnarray}
f_p \propto t^{-47/48} \simeq t^{-0.98} .
\label{eq:fevtk}
\end{eqnarray}
In the thin shell case,
\begin{eqnarray}
R \propto t^{1/5}, \Gamma \propto t^{-2/5}, {\cal N} \propto t^{-6/7}, {\cal E} \propto t^{-8/7}, B \propto t^{-4/7},
\label{eq:gevtn}
\end{eqnarray}
so that
\begin{eqnarray}
\nu_m, \ \nu_c \propto t^{-54/35} \simeq t^{-1.54}
\label{eq:nuevtn}
\end{eqnarray}
and
\begin{eqnarray}
f_p \propto t^{-34/35} \simeq t^{-0.97} .
\label{eq:fevtn}
\end{eqnarray}
Although the evolution before shock crossing ($t<t_\times$) can also be deduced
from suitable expressions for the evolution of $\Gamma$, etc. (e.g. Ioka \& M\'esz\'aros 2004),
it is outside of our interest here.

The RS self-absorption frequency has often been estimated from the blackbody limit approximation
(e.g. Sari \& Piran 1999a,b, Kobayashi \& Zhang 2003),
which, however, suffers from uncertainties in the area of the emitting surface and its time-dependence.
As a unique feature of this work,
we utilize a more proper approach through the appropriate absorption coefficient.
Denoting $\gamma_p=\min(\gamma_m,\gamma_c)$ and $\nu_p=\min(\nu_m,\nu_c)$,
we can use the self-absorption coefficient averaged over an isotropic distribution of pitch angles
(Granot, Piran \& Sari 1999, Panaitescu \& Kumar 2000)
to express the self-absorption optical depth at frequency $\nu_p$ as
\begin{eqnarray}
\tau_p = \psi f(p) e {{\cal N} L \over B \gamma_p^5} \ ,
\label{eq:taupx}
\end{eqnarray}
where $f(p)=(p+2)(p-1)/(3p+2)$, $\psi=2^{8/3} \pi^{5/2}/5 \Gamma(5/6)$,
and $L$ is the comoving frame width of the shocked ejecta.
The expressions for $\gamma_m$ and $\gamma_c$ are
\begin{eqnarray}
\gamma_m = \epsilon_e {p-2 \over p-1} {m_p \over m_e} {\Gamma_0 \over \Gamma}
\label{eq:gmx}
\end{eqnarray}
and
\begin{eqnarray}
\gamma_c = {3 \over 16 \sigma_T c} {m_p \over m_e} {1+z \over \epsilon_B n \Gamma^3 t} .
\label{eq:gcx}
\end{eqnarray}
At $t=t_\times$,
the following relations for ${\cal N}$ and $L$
hold approximately in either the thick and thin cases,
\begin{eqnarray}
{\cal N}_\times = {8 \Gamma_\times^3 \over \Gamma_0} n
\label{eq:nx}
\end{eqnarray}
and
\begin{eqnarray}
L_\times = {\Gamma_\times c t_\times \over 2 (1+z)}
\label{eq:lx}
\end{eqnarray}
(Sari \& Piran 1995, Kobayashi 2000).
Thus, we can evaluate the self-absorption optical depth $\tau_{p,\times}$ at $\nu_p$
at the shock crossing time 
by substituting $t=t_\times$ and $\Gamma=\Gamma_\times$ in Eqs.\ref{eq:taupx}-\ref{eq:gcx}
along with Eqs.\ref{eq:nx} and \ref{eq:lx}.
In numerical terms:\\
when $\gamma_c<\gamma_m$ (fast cooling),
\begin{eqnarray}
\tau_{p,\times} = 2.5 \times 10^{-7} f(p) (1+Y)^5 \epsilon_{B,-2}^{9/2} E_{53}^{9/4} n^{13/4} \Gamma_{0,2}^{-1} T_{2}^{-3/4}
\label{eq:tauftk}
\end{eqnarray}
for the thick shell case, and
\begin{eqnarray}
\tau_{p,\times} = 1.5 \times 10^{-7} f(p) (1+Y)^5 \epsilon_{B,-2}^{9/2} E_{53}^2 n^{7/2} \Gamma_{0,2}
\label{eq:tauftn}
\end{eqnarray}
for the thin shell case;\\
when $\gamma_m<\gamma_c$ (slow cooling),
\begin{eqnarray}
\tau_{p,\times} = 0.34 f(p) \left(p-1 \over p-2\right)^5 \epsilon_{e,-1}^{-5} \epsilon_{B,-2}^{-1/2} E_{53} n^{-1/2} \Gamma_{0,2}^{-6} T_{2}^{-2}
\label{eq:taustk}
\end{eqnarray}
for the thick shell case, and
\begin{eqnarray}
\tau_{p,\times} = 0.91 f(p) \left(p-1 \over p-2\right)^5 \epsilon_{e,-1}^{-5} \epsilon_{B,-2}^{-1/2} E_{53}^{1/3} n^{1/6} \Gamma_{0,2}^{-2/3}
\label{eq:taustn}
\end{eqnarray}
for the thin shell case.

The time evolution of $\tau_p$ after shock crossing can be derived 
in a similar way to Eqs.\ref{eq:gevtk}-\ref{eq:fevtn},
using $B \propto {\cal E}^{1/2}$, $L \propto {\cal N}^{-1} r^{-2}$ and $\tau_p \propto {\cal N} L B^{-1} \gamma_p^{-5}$.
In the thick shell case,
\begin{eqnarray}
{\cal N} \propto t^{-13/16}, L \propto t^{9/16}, B \propto t^{-13/24}, \gamma_p \propto t^{-13/48} ,
\label{eq:nevtk}
\end{eqnarray}
so that
\begin{eqnarray}
\tau_p \propto t^{79/48} \simeq t^{1.65} ,
\label{eq:tevtk}
\end{eqnarray}
whereas in the thin shell case,
\begin{eqnarray}
{\cal N} \propto t^{-6/7}, L \propto t^{16/35}, B \propto t^{-4/7}, \gamma_p \propto t^{-2/7} ,
\label{eq:nevtn}
\end{eqnarray}
so that
\begin{eqnarray}
\tau_p \propto t^{8/5} = t^{1.6} .
\label{eq:tevtn}
\end{eqnarray}
Thus we have the value of $\tau_p$ at any $t \ge t_\times$.

Writing $\nu_b=\max(\nu_m,\nu_c)$,
the self-absorption optical depth at an arbitrary frequency and time after shock crossing is then
\begin{eqnarray}
\tau_\nu &=& \tau_p
\left\{
\begin{array}{ll}
\displaystyle{\left(\nu \over \nu_p\right)^{-5/3}} & \nu < \nu_p \\
\displaystyle{\left(\nu \over \nu_p\right)^{-(q+4)/2}} & \nu_p < \nu < \nu_b \\ 
\displaystyle{\left(\nu \over \nu_b\right)^{-(p+5)/2} \left(\nu_b \over \nu_p\right)^{-(q+4)/2}} & \nu_b < \nu \\
\end{array}
\right. \
\label{eq:taunu}
\end{eqnarray}
The frequency at which $\tau_\nu=1$ gives the synchrotron self absorption frequency $\nu_a$,
\begin{eqnarray}
\nu_a &=& \nu_p
\left\{
\begin{array}{ll}
\displaystyle{\tau_p^{3/5}} & \tau_p < 1 \\
\displaystyle{\tau_p^{2/(q+4)}} & \tau_b < 1 < \tau_p \\ 
\displaystyle{\tau_p^{2/(p+5)} \left(\nu_b \over \nu_p\right)^{1-(q+4)/(p+5)}} & \tau_b > 1\\
\end{array}
\right. \ ,
\label{eq:nua}
\end{eqnarray}
where $\tau_b=\tau_p (\nu_b/\nu_p)^{-(q+4)/2}$ is the optical depth at $\nu_b$.
The case of $\tau_b>1$ can be important for high ambient densities.

The power-law index of the emission spectrum between $\nu_p$ and $\nu_b$ is $(q-1)/2$,
where $q$ is the corresponding index in the electron distribution,
being $q=2$ for fast cooling and $q=p$ for slow cooling.
For the broadband spectrum $f_\nu$,
three cases can be distinguished depending on the location of $\nu_a$ with respect to $\nu_p$ and $\nu_b$:
for $\nu_a < \nu_p$,
\begin{eqnarray}
f_\nu &=& f_p
\left\{
\begin{array}{ll}
\displaystyle{\left(\nu \over \nu_a\right)^2 \left(\nu_a \over \nu_p\right)^{1/3}} & \nu < \nu_a \\
\displaystyle{\left(\nu \over \nu_p\right)^{1/3}} & \nu_a < \nu < \nu_p \\
\displaystyle{\left(\nu \over \nu_p\right)^{-(q-1)/2}} & \nu_p < \nu < \nu_b\\
\displaystyle{\left(\nu \over \nu_b\right)^{-p/2} \left(\nu_b \over \nu_p\right)^{-(q-1)/2}} & \nu_b < \nu\\
\end{array}
\right. \ ,
\label{eq:fnu1}
\end{eqnarray}
for $\nu_p< \nu_a < \nu_b$,
\begin{eqnarray}
f_\nu &=& f_p
\left\{
\begin{array}{ll}
\displaystyle{\left(\nu \over \nu_p\right)^2 \left(\nu_p \over \nu_a\right)^{5/2}} \left(\nu_a \over \nu_p\right)^{-(q-1)/2} & \nu < \nu_p \\
\displaystyle{\left(\nu \over \nu_a\right)^{5/2} \left(\nu_a \over \nu_p\right)^{-(q-1)/2}} & \nu_p < \nu < \nu_a \\
\displaystyle{\left(\nu \over \nu_p\right)^{-(q-1)/2}} & \nu_a < \nu < \nu_b\\
\displaystyle{\left(\nu \over \nu_b\right)^{-p/2} \left(\nu_b \over \nu_p\right)^{-(q-1)/2}} & \nu_b < \nu\\
\end{array}
\right. \ ,
\label{eq:fnu2}
\end{eqnarray}
and for $\nu_b < \nu_a$,
\begin{eqnarray}
f_\nu &=& f_p
\left\{
\begin{array}{ll}
\displaystyle{\left(\nu \over \nu_p\right)^2 \left(\nu_p \over \nu_a\right)^{5/2} \left(\nu_a \over \nu_b\right)^{-p/2}}\\
 \ \ \ \ \ \times \displaystyle{\left(\nu_b \over \nu_p\right)^{-(q-1)/2}} & \nu < \nu_p \\
\displaystyle{\left(\nu \over \nu_a\right)^{5/2} \left(\nu_a \over \nu_b\right)^{-p/2} \left(\nu_b \over \nu_p\right)^{-(q-1)/2}} & \nu_p < \nu < \nu_a\\
\displaystyle{\left(\nu \over \nu_b\right)^{-p/2} \left(\nu_b \over \nu_p\right)^{-(q-1)/2}} & \nu_a < \nu\\
\end{array}
\right. \ .
\label{eq:fnu3}
\end{eqnarray}

\begin{figure}
\centering
\epsfig{file=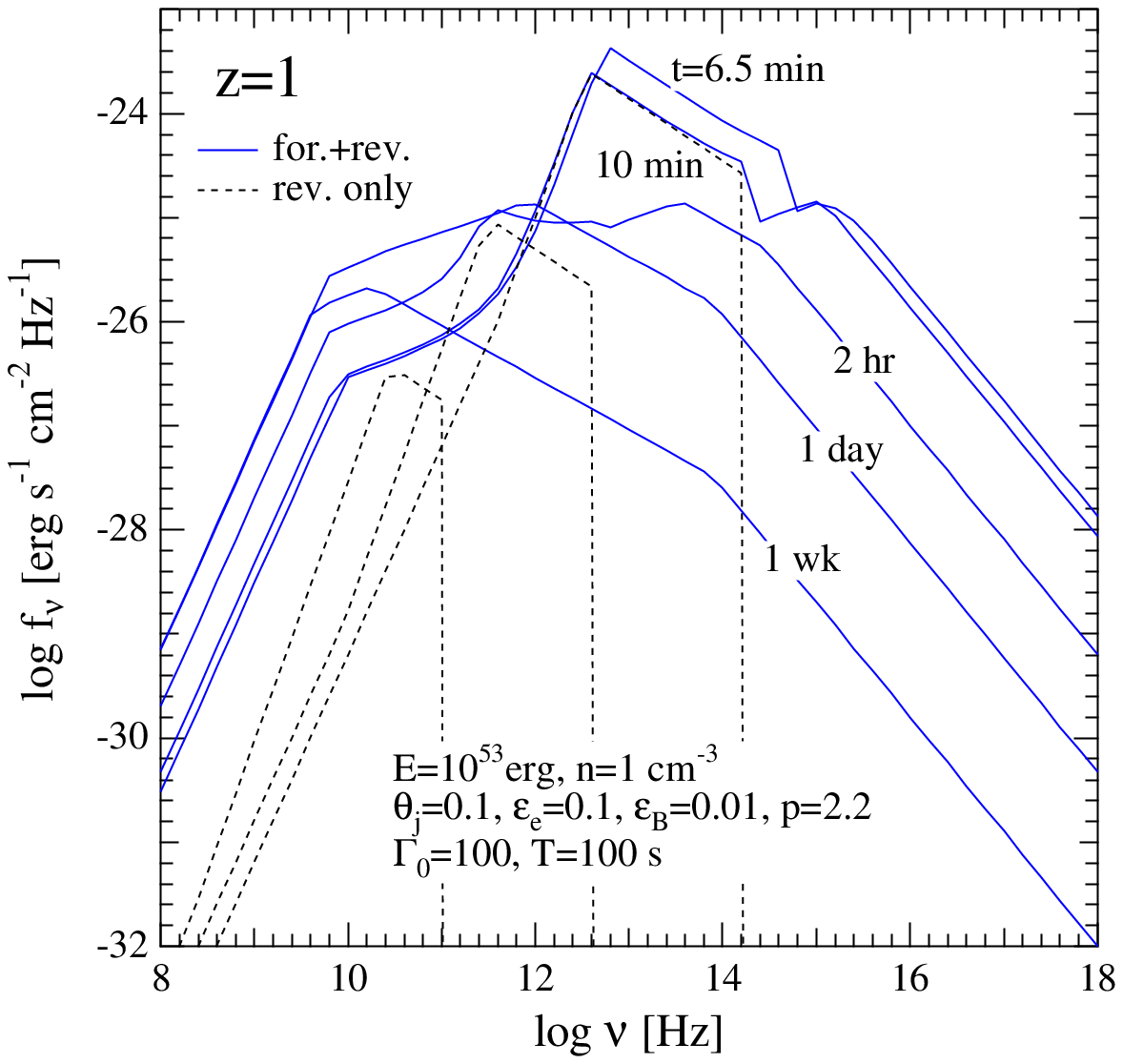,width=0.50\textwidth}
\caption{
Broadband spectra of a GRB afterglow at $z=1$ at selected times after the burst,
for the fiducial parameters discussed in \S \ref{sec:aft}.
The solid curves are the total forward shock plus reverse shock spectra,
while the dashed curves show the reverse shock component only for $t=$10 min, 2 hr and 1 day.
}
\label{fig:sz1}
\end{figure}

Eqs.\ref{eq:numrel}-\ref{eq:fnu3} give a complete description
of the self-absorbed RS spectrum at any time $t$ after shock crossing.
Employing this formulation,
Fig.\ref{fig:sz1} shows the time-dependent, broadband afterglow spectra of a burst at $z=1$
with our fiducial parameters discussed in \S \ref{sec:aft},
where the RS contribution has been displayed separately along with the total FS plus RS emission.

\label{lastpage}

\end{document}